\documentclass[aps,prd,twocolumn,superscriptaddress,groupedaddress,footnote,nofootinbib]{revtex4}
\usepackage{graphicx}  
\usepackage{dcolumn}   
\usepackage{bm}        
\usepackage{amssymb}   
\usepackage{amsmath}
\usepackage{xcolor}
\usepackage{rotating}
\usepackage{amsthm}
\usepackage{subfig}
\usepackage{subfloat}
\usepackage{actuarialsymbol}
\usepackage{braket}
\usepackage{mathtools}
\usepackage{bbold}
\usepackage{amssymb}
\usepackage{enumitem}
\usepackage{comment}
\usepackage[colorlinks=true, citecolor=blue, urlcolor=blue]{hyperref}  
\usepackage{MnSymbol}
\usepackage{cancel}

\newcommand{\be}{\begin{equation}}\newcommand{\ee}{\end{equation}}
\newcommand{\bea}{\begin{eqnarray}}\newcommand{\eea}{\end{eqnarray}}
\newcommand{\brr}{\begin{array}}\newcommand{\err}{\end{array}}
\newcommand{\bit}{\begin{itemize}}\newcommand{\eit}{\end{itemize}}
\newcommand{\ben}{\begin{enumerate}}\newcommand{\een}{\end{enumerate}}

\newcommand{\ba}{\begin{array}}
\newcommand{\ea}{\end{array}}

\definecolor{darkred}{rgb}{.8,0,0}

\definecolor{darkblue}{rgb}{0,0,.7}

\def\1{{_{1}}}\def\2{{_{2}}}

\def\noHe0{:\;\!\!\;\!\!:H_e(0):\;\!\!\;\!\!:}
\def\noHm0{:\;\!\!\;\!\!:H_\mu(0):\;\!\!\;\!\!:}

\def\1{{_{1}}}\def\2{{_{2}}}

\usepackage{ragged2e}

\begin{document}
\title{Characterizing entanglement dynamics in QED scattering processes}

\author{M. Blasone}
\email{blasone@sa.infn.it}
\affiliation{Dipartimento di Fisica, Universit\`a di Salerno, Via Giovanni Paolo II, 132 I-84084 Fisciano (SA), Italy}
\affiliation{INFN, Sezione di Napoli, Gruppo collegato di Salerno, Italy}

\author{S. De Siena}
\email{silvio.desiena@gmail.com}
\affiliation{(Ret. Prof.) Universit\`a di Salerno, Via Giovanni Paolo II, 132 I-84084 Fisciano (SA), Italy}

\author{G. Lambiase}
\email{lambiase@sa.infn.it}
\affiliation{Dipartimento di Fisica, Universit\`a di Salerno, Via Giovanni Paolo II, 132 I-84084 Fisciano (SA), Italy}
\affiliation{INFN, Sezione di Napoli, Gruppo collegato di Salerno, Italy}

\author{B. Micciola}
\email{bmicciola@unisa.it}
\affiliation{Dipartimento di Fisica, Universit\`a di Salerno, Via Giovanni Paolo II, 132 I-84084 Fisciano (SA), Italy}
\affiliation{INFN, Sezione di Napoli, Gruppo collegato di Salerno, Italy}
\affiliation{Fakult\"{a}t f\"{u}r Mathematik, Universit\"{a}t Wien, Oskar-Morgenstern-Platz 1, 1090 Vienna, Austria}

\author{K. Simonov}
\email{kyrylo.simonov@univie.ac.at}
\affiliation{Fakult\"{a}t f\"{u}r Mathematik, Universit\"{a}t Wien, Oskar-Morgenstern-Platz 1, 1090 Vienna, Austria}

\date{\today}

\begin{abstract}

We study the entanglement among helicity degrees of freedom in quantum electrodynamics (QED) scattering processes at fixed momentum, modeled as quantum maps whose spectral structure determines the associated entanglement dynamics. We show that, for fermion-fermion scatterings, parity invariance preserves maximal entanglement and generically leads to entanglement saturation under iterations of the maps. The structure of the maps  also provides an accurate description of the entanglement dynamics in processes involving both fermions and photons.

\end{abstract}

\maketitle

\emph{Introduction}. The physics of fundamental interactions among elementary particles and the concepts and tools of quantum information theory have begun to intersect in recent years \cite{Bertlmann2006, DiDomenico:2007zza, Blasone:2023qqf, Shi:2025fiz, Mavromatos:2025mxa}\footnote{A prominent example is provided by the role of quantum entanglement in the experimental discovery of the Higgs boson at CERN in 2012.}. More generally, the application of quantum information methods to fundamental interactions offers new avenues for probing new physics beyond the Standard Model and for testing basic principles of quantum mechanics \cite{DiDomenico:1995ky, Bertlmann:2001ea, Blasone:2007vw, Banerjee:2015mha, Formaggio:2016cuh, Beane:2018oxh, Dixit:2018gjc, Ming:2020nyc, Wang:2020vdm, Low:2021ufv, Blasone:2021cau, Bittencourt:2022tcl, Bittencourt:2023asd, Low:2024hvn, Kowalska:2024kbs, Duch:2024pwm, Carena:2025wyh, Chang:2024wrx, Liu:2025pny, Martin:2025hzm, Nunez:2025dch, Thaler:2024anb, Liu:2025bgw, Liu:2025qfl, McGinnis:2025brt, McGinnis:2025xgt}. Conversely, the study of entanglement generated in high-energy processes may also contribute to the development of novel and effective quantum-information protocols. The identification of experimental methods capable of revealing entanglement generated in fundamental interaction \cite{Fabbrichesi:2021npl, Barr:2021zcp, Afik:2022kwm, HZZ, Afik:2023dgh, LHC, QIHEP, CMS, ATLAS:2023fsd,Maltoni:2024tul, Barr:2024djo, Altomonte:2024upf, Gao:2025kdi, Jaloum:2025bkx}, with significant progress for entanglement involving top quarks~\cite{CMS, ATLAS:2023fsd} and $Z$-bosons~\cite{ATLAS:2026hye}, has already been achieved. Recently, a quantum process tomography approach for scattering processes has been developed in Ref.~\cite{Altomonte:2024upf} in terms of quantum instruments and Choi matrices. At the same time, extensive theoretical efforts are devoted to understanding the generation and evolution of entanglement in a variety of fundamental interactions~\cite{Seki:2014cgq, Peschanski:2016hgk, Cervera-Lierta:2017tdt, Beane:2018oxh, Araujo, Fan, Fan:2017mth, Fonseca:2021uhd, Sinha:2022crx, Serafini, Blasone:2024dud,Fedida:2024dwc, BlasoneSE2, Cao:2026mza, Cao:2026ljz, Xie:2026}.

The above studies have also impact on new technological developments as quantum lithography and quantum positron emission tomography (QE-PET), the latter exploiting the strong entanglement between photons produced in pair annihilation \cite{Boto2000, Toghyani2016, Watts2021, Moskal2025}. This reflects an ongoing shift toward a new generation of devices based on QED rather than quantum optics, for which the corresponding theoretical tools are currently under development \cite{Smeets:2025wro, Asenov:2025ueu,Clarke2026}.

Refs.~\cite{Cervera-Lierta:2017tdt, Serafini, Blasone:2024dud, BlasoneSE2} investigated helicity entanglement in QED scattering processes by considering scattering at fixed momenta, i.e. a generalized measurement described by a positive operator-valued measure, which results in a classical probability and a post-measurement state. In this context, entanglement can be generated starting from simple separable states~\cite{Cervera-Lierta:2017tdt, Serafini}. In Ref.~\cite{BlasoneSE2}, a detailed study of entanglement dynamics in QED for generic initial (pure) states was performed by means of the complete complementarity relations, which characterize both local and non-local properties of a quantum state.

In this Letter, we completely characterize the entanglement dynamics in tree--level QED scattering processes, modeling them as (non-linear) quantum maps. When only fermions are involved, we find that  maximal entanglement is  preserved in the scattering, and that the  set of maximally entangled states is an attractor, with a  basin of attraction which progressively increases going to the relativistic regime. We show that these results are a consequence of parity invariance, which shapes the general form of the maps and determines their spectral properties, constraining entanglement dynamics.  
The entanglement saturation emerging in the iteration of the maps appears as a form of irreversibility for  fundamental interaction processes involving elementary particles.
Finally, if processes involve both fermions and photons, we show that entanglement saturation is not achieved, however modeling in terms of maps  provides the fundamental features of the entanglement dynamics.

\vspace{0.3cm}

\emph{Formalism and assumptions}. We briefly summarize the formalism and assumptions adopted in this work, following Refs.~\cite{Cervera-Lierta:2017tdt, Serafini, Blasone:2024dud, BlasoneSE2}. Scattering processes are described by the unitary $S$-matrix evolution of the system from an initial state $\rho^{\mathrm{in}}$ at $t=-\infty$ to a final state 
\begin{equation}
    \rho^f = S \rho^{\mathrm{in}} S^{\dagger} \label{finalS}
\end{equation}
at $t = \infty$. Working in the center-of-mass (COM) frame, we focus on the spin degrees of freedom of the incoming and outgoing particles at fixed momenta.

We assume sharp momentum filtering of the outgoing particles without resolving their helicities. This amounts to post-selecting a fixed outgoing momentum configuration $q$, described by the projector
\begin{equation}
    \Pi_q = \sum_{r,s\in\{R,L\}} \ket{q;rs}\bra{q;rs},
\end{equation}
where $R$ and $L$ denote right- and left-handed helicity states, respectively. The corresponding normalized conditional state is
\begin{equation}
    \tilde{\rho}^f = \frac{\Pi_q \rho^f \Pi_q}{\operatorname{Tr}[\Pi_q \rho^f \Pi_q]}. \label{PMFS-gen}
\end{equation}
Expanding this state in the outgoing helicity basis gives
\begin{equation}
    \tilde{\rho}^f =  \sum_{r,s,r',s'}
    \tilde{\rho}^{f}_{rs;r's'}
    \ket{q;rs}\bra{q;r's'}. \label{PMS}
\end{equation}
The matrix elements $\tilde{\rho}^{f}_{rs; r's'}$ can be expressed in terms of the $S$-matrix elements $S^{(qp)}_{rs; ij} = \bra{q, rs} S\ket{p, ij}$. The latter are related to the scattering amplitudes $\mathcal{M}_{rs;ij}$ via 
\begin{equation}\label{def:ampiezze}
    S^{(qp)}_{rs; ij} = i (2\pi)^4 \delta^{(4)}(p-q) \mathcal{M}_{rs; ij},   
\end{equation}
Here and throughout, the first pair of helicity indices labels the outgoing state, whereas the second pair labels the incoming state. For fixed incoming and outgoing momenta, the amplitudes $\mathcal M_{rs;ij}$ therefore fully characterize the states of the outgoing particles after scattering (see Ref.~\cite{Serafini} for a detailed discussion).

For a given process, nonlocal helicity correlations can then be analyzed by computing the scattering amplitudes in the COM frame at a given perturbative order. These amplitudes depend on the scattering angle $\theta$ and on the dimensionless parameter $\mu=|\vec{p}|/m$ ($m$ is the electron mass), which specifies the kinematical regime. In what follows,  scattering amplitudes are computed at tree level\footnote{As argued in Ref.~\cite{Serafini}, loop corrections do not substantially affect entanglement production.}.

\vspace{0.3cm}

\emph{Scattering quantum maps}. 
We now formulate the momentum-conditioned scattering process as an operation on the helicity density matrix. We use the ordered helicity basis $\{ \ket{RR},\ket{RL},\ket{LR},\ket{LL}\}$ and introduce the matrices $\mathbf{M}$ defined in terms of the scattering amplitudes as 
\begin{equation}
\mathbf M =
\begin{pmatrix}
\mathcal M_{RR;RR} & \mathcal M_{RR;RL} &
\mathcal M_{RR;LR} & \mathcal M_{RR;LL} \\[0.2em]
\mathcal M_{RL;RR} & \mathcal M_{RL;RL} &
\mathcal M_{RL;LR} & \mathcal M_{RL;LL} \\[0.2em]
\mathcal M_{LR;RR} & \mathcal M_{LR;RL} &
\mathcal M_{LR;LR} & \mathcal M_{LR;LL} \\[0.2em]
\mathcal M_{LL;RR} & \mathcal M_{LL;RL} &
\mathcal M_{LL;LR} & \mathcal M_{LL;LL}
\end{pmatrix},
\label{SAM}
\end{equation}
For the tree-level QED processes considered below, the amplitudes can be chosen real in the conventions adopted here, so that $\mathbf{M}^\dagger = \mathbf{M}^T$.

The incoming helicity state at fixed momentum $p$ is given by
\begin{align}
    \rho^{\mathrm{in}} &= \sum_{i,j,k,l} \, \rho^{\mathrm{in}}_{ij; kl} \, \ket{p; ij}\bra{p; kl},
    \label{IMS}
\end{align}
where $\rho^{\mathrm{in}}_{ij;kl} = \bra{p;ij}\rho^{\mathrm{in}}\ket{p;kl}$. Using Eqs.~\eqref{finalS} and \eqref{PMFS-gen}, the matrix elements of the momentum-conditioned state are
\begin{align}
    {}\hspace{-2.3mm}
    \tilde{\rho}^{f}_{rs; r's'} = \frac{\sum_{i, j, k, l} \rho^{\mathrm{in}}_{ij; kl} \bra{q; rs} S \ket{p; ij} \bra{p; kl} S^{\dagger}\ket{q; r's'}}{\sum_{r,s, i, j, k, l} \rho^{\mathrm{in}}_{ij; kl} \bra{q; rs} S \ket{p; ij} \bra{p; kl} S^{\dagger}\ket{q; rs}}.
    \label{PMFS1}
\end{align}
Expressing the $S$-matrix elements in terms of the scattering amplitudes via Eq. \eqref{def:ampiezze}, we obtain
\begin{equation}
    \tilde{\rho}^{f}_{rs; r's'} = \frac{\sum_{i, j, k, l} \mathcal{M}_{rs; ij} \rho^{\mathrm{in}}_{ij; kl} \mathcal{M}_{r's'; kl}}{\sum_{i, j, k, l, r, s} \mathcal{M}_{rs; ij} \rho^{\mathrm{in}}_{ij; kl} \mathcal{M}_{rs; kl}},
\end{equation}
or, in a compact form,
\bea
    \tilde{\rho}^f &=& \frac{\textbf{M} \rho^{\rm in} \textbf{M}^T}{\operatorname{Tr} [\textbf{M} \rho^{\rm in} \textbf{M}^T]}.
\label{PMFS2}
\eea
Eq.~\eqref{PMFS2} defines a quantum map, namely a transformation that associates an outgoing quantum state with a given incoming state. In quantum information theory, physical transformations of density matrices are described by completely positive (CP) maps, which preserve the positivity of quantum states even when the system is entangled with additional degrees of freedom. Deterministic transformations are also trace-preserving (CPTP), whereas probabilistic transformations associated with selected measurement outcomes are trace-nonincreasing.

In our case, the map describes how the helicity state changes during scattering, conditioned on a fixed outgoing momentum configuration. It admits a natural interpretation in terms of quantum instruments, namely collections of quantum maps describing both the probabilities of different measurement outcomes and the corresponding changes in the quantum state\footnote{For background on quantum channels, quantum operations, and quantum instruments, including their representations in terms of Choi matrices, see, e.g., Refs.~\cite{Nielsen:2012yss, Bassano, Altomonte:2024upf, Homa:2024mzd}.}. While momentum postselection is projective on the full Hilbert space, it induces a generalized measurement on the incoming helicity degrees of freedom. The corresponding CP, trace-nonincreasing operation is $\mathcal E_q[\rho] = K_q\rho K_q^\dagger$, where the Kraus operator $K_q$ is proportional to $\mathbf{M}$, up to kinematic normalization factors. The associated POVM element $F_q = K_q^\dagger K_q$ determines the probability of the selected outcome, $\operatorname{Tr}[\rho^{\mathrm{in}}F_q]$. The kinematic factors cancel upon normalization, yielding Eq.~\eqref{PMFS2}.

We now consider repeated applications of the same scattering map at fixed kinematic parameters. After $n$ iterations,
\begin{equation}
    \tilde{\rho}_n = \mathcal{N}_n^{-1} \mathbf{M}^n\rho^{\mathrm{in}} (\mathbf{M}^T)^n ,
    \label{ItMS}
\end{equation}
where
\begin{equation}
    \mathcal{N}_n = \operatorname{Tr}\!\left[ \mathbf{M}^n\rho^{\mathrm{in}} (\mathbf{M}^T)^n \right]
\end{equation}
denotes the normalization of the conditional state obtained after $n$ successive scatterings. 

If the initial helicity state is pure,
$\rho^{\mathrm{in}} = \ket{i}\bra{i}$, the conditional state remains
pure. After one scattering event,
\begin{equation}
    \ket{f} = \frac{\mathbf{M}\ket{i}}{\lVert \mathbf{M}\ket{i}\rVert},
    \label{PMSPS}
\end{equation}
whereas after $n$ iterations,
\begin{equation}
    \ket{f_n} = \frac{\mathbf{M}^n\ket{i}}{\lVert \mathbf{M}^n\ket{i}\rVert},
    \label{ItPS}
\end{equation}
where $\lVert \ket{\psi} \rVert = \sqrt{\langle\psi|\psi\rangle}$ is the standard norm on the Hilbert space spanned by helicity states.
\vspace{0.3cm}

\emph{The role of parity invariance}.
For $2 \to 2$ scattering processes, parity invariance imposes nontrivial constraints on the helicity amplitudes~\cite{JacobWick}. For spin-$\tfrac{1}{2}$ particles, these constraints reduce the matrix~(\ref{SAM}) to the form\footnote{For a more detailed treatment of this section, see Appendix~\ref{app:parity}.}
\begin{equation}\label{genmappar}
    \mathbf{M} = \left(
    \begin{array}{cccc}
        A & B & C & D \\
        H & E & F & - I \\
        I & F & E & - H \\
        D & -C & -B & A
    \end{array}
    \right),
\end{equation}
where $A=\mathcal{M}_{RR;RR}$, etc., according to Eq.~\eqref{SAM}.
The matrix \eqref{genmappar} exhibits the following properties rooted in the symmetries of the theory:
\begin{enumerate}
    \item \emph{The set of maximally entangled states is invariant under the map \eqref{PMFS2}};
    \item \emph{The eigenvectors of $\mathbf{M}$ are maximally entangled states, or special linear combinations of Bell states;}
    \item \emph{For any $n$, the power $\mathbf{M}^n$ has the same form as $\mathbf{M}$.}
\end{enumerate}
The first of these properties was originally established in Ref.~\cite{NoiChaos}. As shown below, all three properties follow naturally from the representation of $\mathbf{M}$ in the  ordered Bell basis $\{\ket{\Phi^+}, \ket{\Psi^-}, \ket{\Phi^-}, \ket{\Psi^+}\}$, where 
\begin{equation}
    \ket{\Phi^{\pm}}  = \tfrac{1}{\sqrt{2}} (\ket{RR} \pm \ket{LL}), \quad \ket{\Psi^{\pm}} = \tfrac{1}{\sqrt{2}} (\ket{RL} \pm \ket{LR}).
\end{equation}
In this basis, the matrix \eqref{genmappar} takes the block-diagonal form
\bea
    \label{MBell}
    \mathbf{M}_{\rm Bell} &=&
    \left(
    \begin{array}{cc}
        R_1 & 0 \\
        0 & R_2
    \end{array}
    \right), \;\; R_{1,2}=\left(
    \begin{array}{cc}
        A\pm D & B\mp C  \\
        H\mp I & E\mp F 
    \end{array}
    \right).
\eea
This decomposition identifies the two invariant subspaces $\mathcal{H}_1 = \mathrm{span}\{\ket{\Phi^+}, \ket{\Psi^-}\}$ and $\mathcal{H}_2 = \mathrm{span}\{\ket{\Phi^-}, \ket{\Psi^+}\}$. 

The block decomposition immediately yields a simple proof that the map \eqref{PMFS2} preserves maximal entanglement. Indeed, every maximally entangled two-qubit state can be written in the \emph{magic basis} as~\cite{Lamoureux:2003apz}
\begin{equation}\label{mes}
    \ket{\psi_{\rm ME}} = x_1\ket{\Phi^+} + x_2\ket{\Psi^-} + i x_3\ket{\Phi^-} + i x_4\ket{\Psi^+},
\end{equation} 
where $x_j\in\mathbb{R}$. Applying Eq.~\eqref{MBell} gives
\begin{align}
    \!\!\mathbf{M}_{\rm Bell}\ket{\psi_{\rm ME}} &= R_1 \left(
    \begin{array}{c}
        x_1\\
        x_2
    \end{array}
    \right) + i R_2 \left(
    \begin{array}{c}
        x_3\\
        x_4
    \end{array}
    \right) \nonumber \\
    &= y_1\ket{\Phi^+} + y_2\ket{\Psi^-} + i y_3\ket{\Phi^-} + i y_4\ket{\Psi^+},
\end{align}
with $y_j\in\mathbb{R}$, which is again of the form~\eqref{mes}. Hence the map \eqref{PMFS2} preserves the set of maximally entangled states.

We now consider  iterations of  the map \eqref{PMFS2} and asymptotic convergence. The block structure \eqref{MBell} determines the asymptotic behavior for repeated applications of the map~\eqref{PMFS2}.

Let $\{\ket{\lambda_s}\}_{s=1,\dots,4}$ denote the normalized eigenvectors of $\mathbf{M}$, associated to eigenvalues $\{\lambda_s\}_s$, which form a (non-orthogonal) basis. A pure initial state $\ket{i}$ can be expanded as $\ket{i} = \sum_s c_s \ket{\lambda_s}$, with suitable coefficients $\{c_s\}_s$. Therefore, Eq.~(\ref{ItPS}) yields the normalized $n$-th iteration
\begin{equation}
        \ket{f_n} \,=\, \frac{\sum_{s} \, c_s \, \lambda_{s}^{n} \, \ket{\lambda_s}}{\lVert \sum_{s} \, c_s \, \lambda_{s}^{n} \, \ket{\lambda_s} \rVert}.
\label{ISEMP}
\end{equation}
Similarly, an arbitrary mixed state \eqref{IMS} evolves as
\begin{equation}
    \!\tilde{\rho}_{n} = \mathcal{N}_{n}^{-1} \sum_{i, j, k, l} \rho^{\mathrm{in}}_{ij; kl}  \sum_{s, s'} c_{s}^{(ij)} c_{s'}^{(kl)*} (\lambda_{s} \lambda_{s'}^*)^{n}\ket{\lambda_s}\bra{\lambda_{s'}}, 
    \label{NitMS}
\end{equation}
due to Eq.~\eqref{ItMS}, where $\ket{p; ij} = \sum_s c_s^{(ij)} \ket{\lambda_s}$.

Eqs.~\eqref{ISEMP}--\eqref{NitMS} show that the behavior at large $n$ is entirely determined by the spectral structure of \eqref{genmappar}. As a consequence of parity invariance, two cases occur: 
\begin{itemize}
     \item \emph{All the eigenvectors are maximally entangled:} If the modulus-dominant eigenvalue is non-degenerate, the iterations converge to the corresponding maximally entangled eigenstate. Otherwise, the asymptotic state is, in general, a superposition of the dominant eigenvectors and is therefore not maximally entangled\footnote{A special case arises when either block $R_1$ or $R_2$ in \eqref{MBell} is non-diagonalizable, see Appendix~\ref{app:dynGen}.}.
     \item \emph{The eigenvectors are separable:} The asymptotic state is in general not maximally entangled. 
\end{itemize}
These two scenarios are analyzed in detail below for the physically relevant case of Bhabha scattering, see also Fig.~\ref{fig:degenerate-regions}.

\smallskip

Finally, the asymptotic rate of convergence is controlled by the spectral gap that plays the role of an effective Lyapunov exponent for the induced dynamics. From Eqs. \eqref{ISEMP}--\eqref{NitMS}, if $\lambda_m$ is the modulus-dominant eigenvalue, and $\lambda_s$ ($s\neq m$) the remaining ones, defining $r_m \doteq \ln|\lambda_m|$ and $r_s \doteq \ln|\lambda_s|$, convergence occurs for $n \gg n_c = \lceil(\min_s 2(r_m-r_s))^{-1}\rceil$, where $\lceil\cdot\rceil$ denotes the ceiling value. The quantities $r_m-r_s$ characterize the exponential suppression of subleading components under interaction, and thus determine the rate at which the asymptotic state is approached.

\smallskip

\emph{QED scattering processes involving fermions only}. 
We consider QED scattering processes involving only fermions, e.g.: $e^{-} e^{+} \rightarrow e^{-} e^{+}$ (Bhabha scattering), $e^{-} e^{-} \rightarrow e^{-} e^{-}$ (M$\o{}$ller scattering), $e^{-} \mu^{-} \rightarrow e^{-} \mu^{-}$ and $e^{-} e^{+} \rightarrow \mu^{-} \mu^{+}$ \footnote{For this last process we do not consider iterations, due to the mismatch between input and output particles, although the associated map formally displays the same properties as those of the other three processes.}. Iterations of the map Eq. \eqref{PMFS2} are performed at fixed scattering angle $\theta$: after the first interaction, the output state at a given angle is selected and used as the input state for the next interaction, and the procedure is repeated by selecting the same angle at each step.

\begin{figure}[t]
    \centering
    \includegraphics[width=\columnwidth]{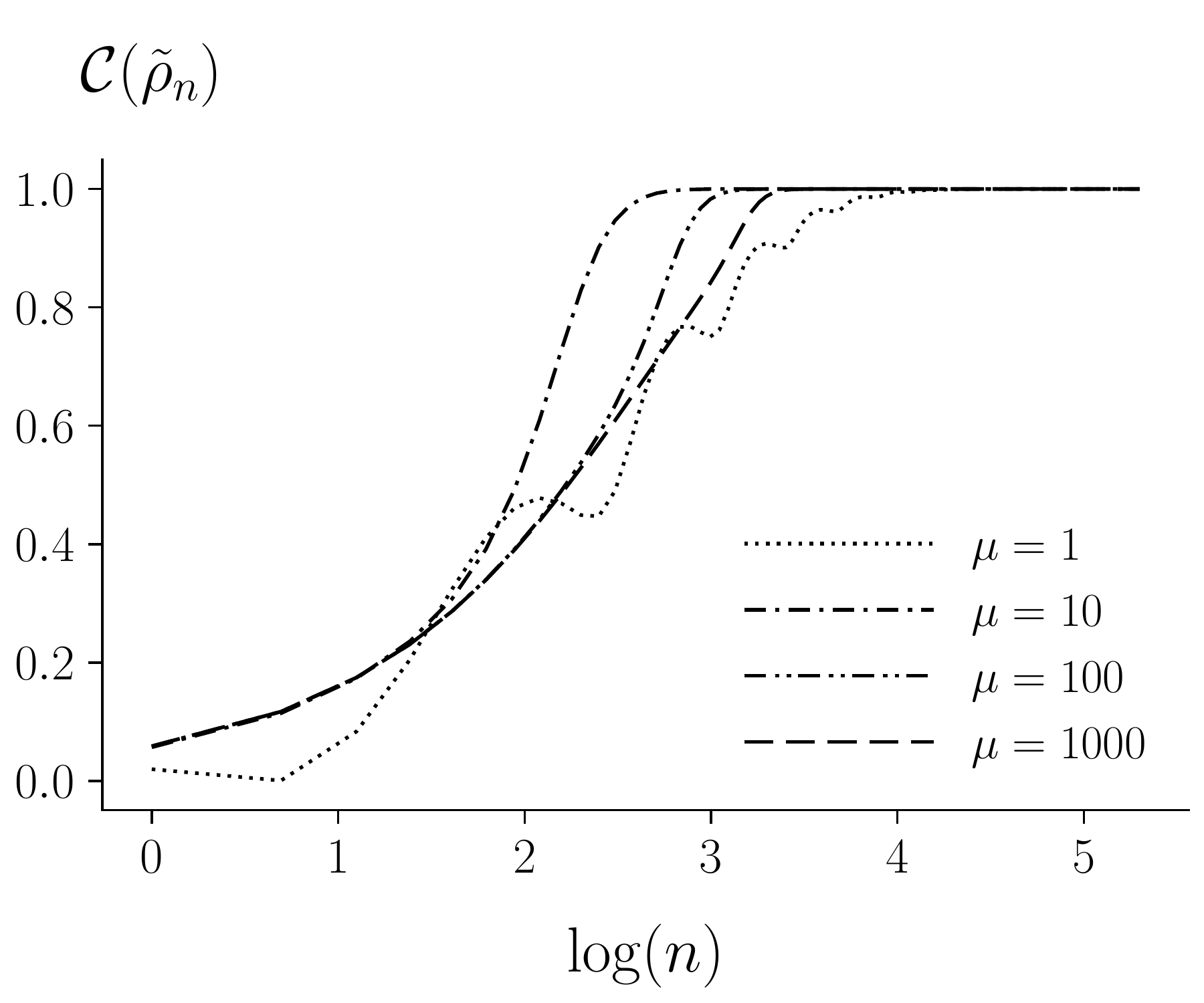}
    \caption{\protect\justifying Concurrence of the post-measurement state $\tilde{\rho}_n$ in the iterated Bhabha process for different regimes, with $\rho^{\rm in } = \ket{RL}\bra{RL}$ and $\theta = \pi/4$.}
    \label{fig1}
\end{figure}

\begin{figure}[t]
    \centering
\includegraphics[width=\columnwidth]{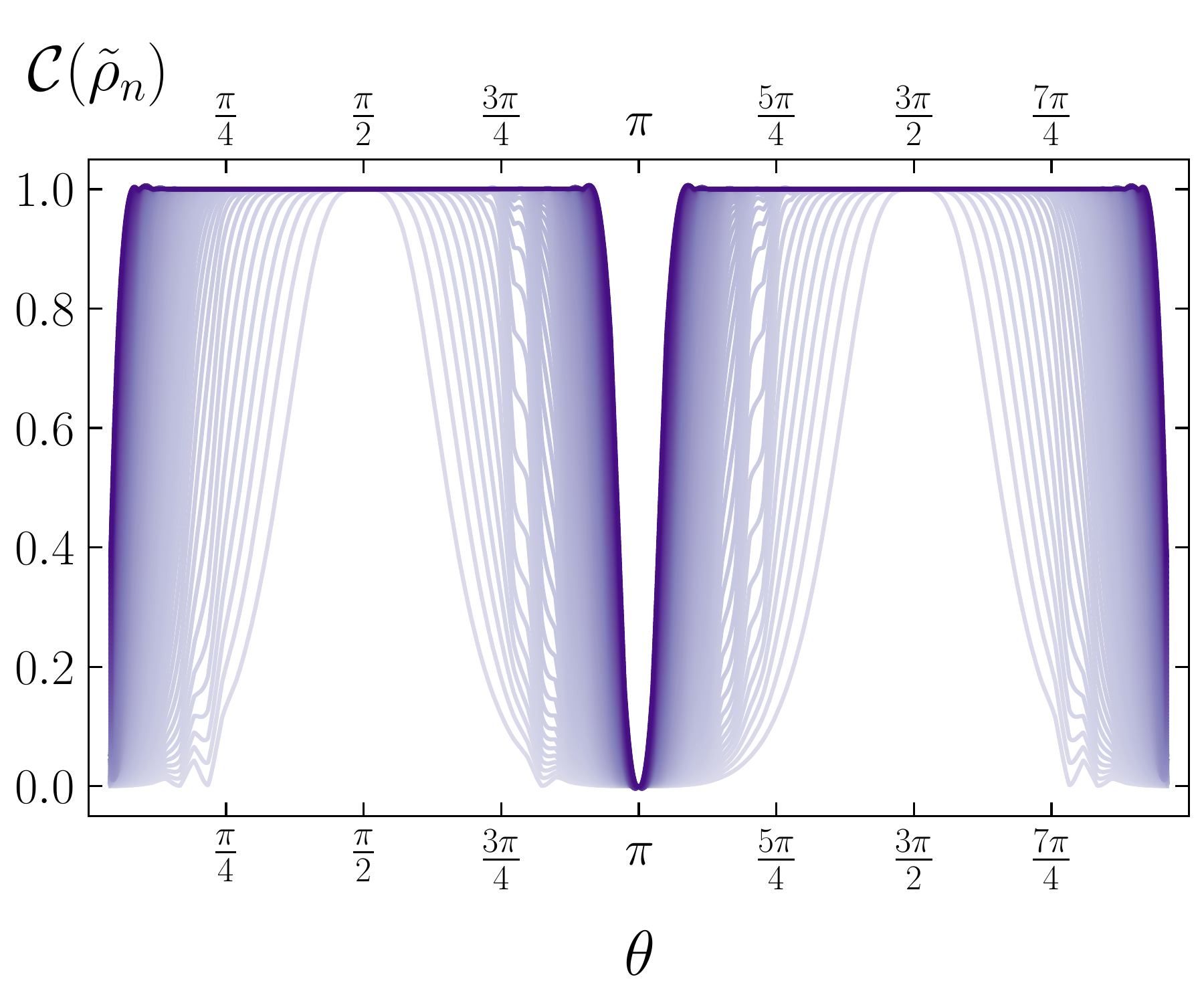}
    \caption{\protect\justifying Concurrence of the post-measurement state $\tilde{\rho}_n$ after $n = 1, \ldots, 100$ iterations of the Bhabha process, with $\rho^{\rm in} = \ket{RL}\bra{RL}$ and $\mu=10$. The dark violet curve corresponds to $n=100$.}
    \label{fig3}
\end{figure}

We quantify entanglement by means of the concurrence, defined for any state $\rho$ as
\begin{equation}
    \mathcal{C}(\rho) = \max\{0, \omega_1-\omega_2-\omega_3-\omega_4\},    
\end{equation}
where $\{\omega_s\}_{s=1,\ldots,4}$ are the eigenvalues, in descending order, of the matrix $\sqrt{\sqrt{\rho}\rho'\sqrt{\rho}}$, with $\rho' = (\sigma_2\otimes\sigma_2)\rho^{\ast}(\sigma_2\otimes\sigma_2)$, and $\sigma_2$ is the second Pauli matrix \cite{Wootters:1997id}. In Fig.~\ref{fig1}, it is shown the saturation of the concurrence, when iterating the map (at a specific angle) starting from the initial pure separable state $|RL\rangle$, for the case of Bhabha scattering. In Fig.~\ref{fig3}, the same iteration process is shown over all scattering angles. In Table~\ref{tab1}, we list the maximally entangled states obtained asymptotically by iterating the M\o{}ller and Bhabha scattering maps with initial states $|RL\rangle$ and $|RR\rangle$. These results reproduce the maximally entangled states reported in the tables of Ref.~\cite{BlasoneSE2}. Note that the existence of entanglophobous states \cite{BlasoneSE2} makes it not obvious \emph{a priori} that one must have a saturation trend. Indeed, for  $\mu = 1$ in Fig.~\ref{fig1}, we observe initial oscillations due to the entanglophobous action which, however, rapidly decay.

\begin{figure}[t]
    \centering
    \includegraphics[width=\columnwidth]{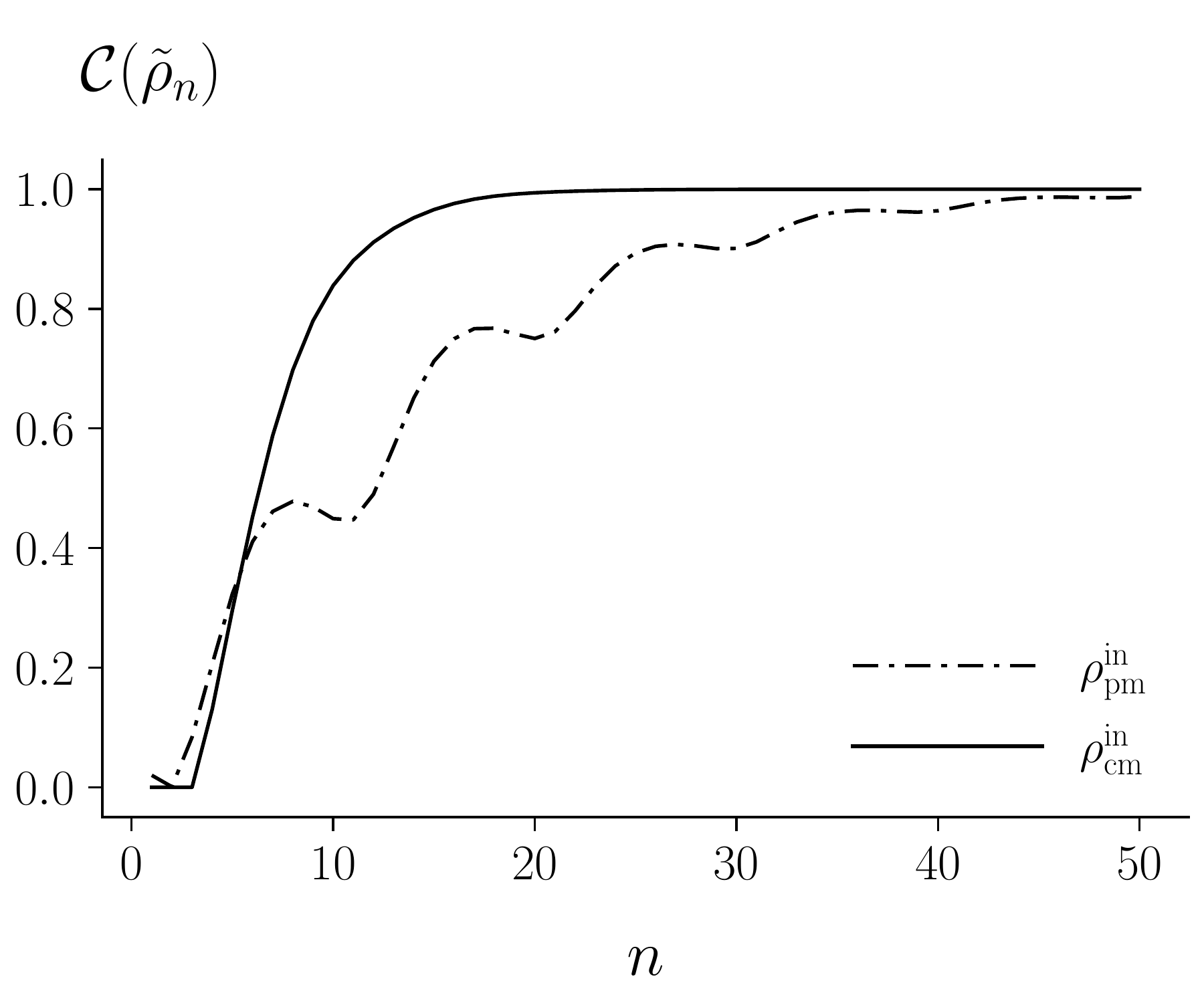}
    \caption{\protect\justifying Concurrence of the post-measurement state $\tilde{\rho}_n$ in the iterated Bhabha process for partially $\rho^{\mathrm{in}}_{\rm pm}$ and completely $\rho^{\mathrm{in}}_{\rm cm}$ mixed initial states, with $\mu = 1$ and $\theta = \pi/4$.}
    \label{fig2}
\end{figure}

\begin{figure*}[t]
    \centering

    \includegraphics[width=\textwidth]{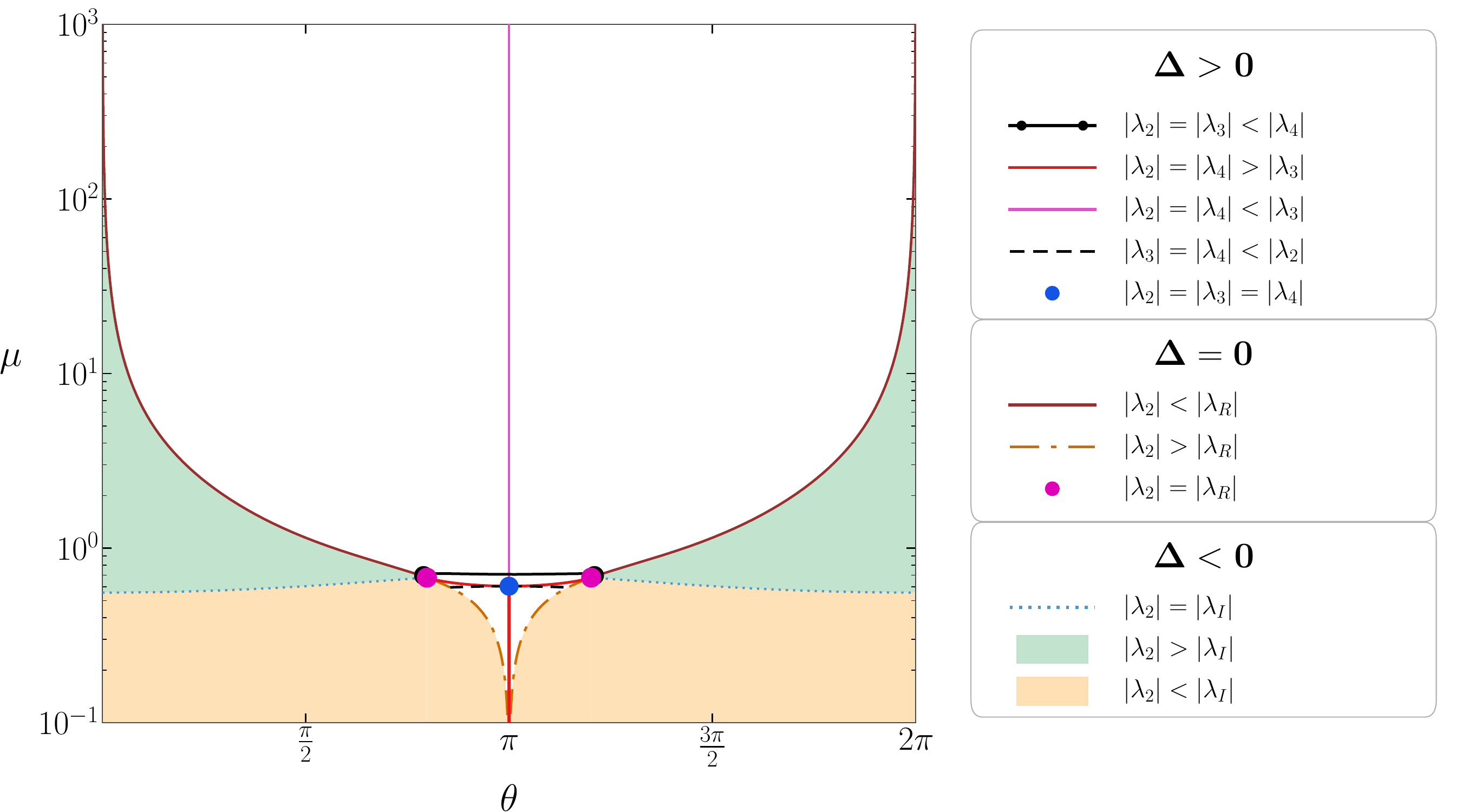}
    \caption{\protect\justifying Regions in the plane ($\theta, \mu$) characterized by the spectral properties of the Bhabha matrix $\mathbf{M}_B$ for the case $c_1=0$  in the expansions (\ref{ISEMP}) or (\ref{NitMS}). For $\Delta > 0$, white region corresponds to  non-degenerate modulus-dominant eigenvalues. In this region, iteration of map \eqref{PMFS2} leads asymptotically to a maximally entangled state, for any initial state. In the presence of modulus degenerate eigenvalues, the asymptotic states are not maximally entangled. These degeneracies are shown as the black, red, pink, and dashed black curves, corresponding to the different relations among the eigenvalue moduli. For $\Delta < 0$, $\lambda_I \doteq \lambda_3 =\lambda_4^*$, and   entanglement saturation is not a priori guaranteed. Finally, for $\Delta = 0$, $\lambda_R \doteq \lambda_3 = \lambda_4$,  the eigenvectors $\ket{\lambda_3}$ and $\ket{\lambda_4}$ coalesce (bordeaux and dashed-dot red curves): this triggers the \textit{exceptional point mechanism} \cite{Kato, Minganti2019, Schaefer2022, Quarteroni} that in the present case leads to entanglement saturation.}
    \label{fig:degenerate-regions}
\end{figure*}

Finally, Fig.~\ref{fig2} illustrates complete saturation starting from the partially mixed state $\rho^{\mathrm{in}}_{\rm pm} = \frac{1}{2}\bigl( \ket{RL}\bra{RL} + \ket{LR}\bra{LR} \bigr)$ and the completely mixed state $\rho^{\mathrm{in}}_{\rm cm} = \frac{1}{4} \sum_{s,l \in \{R,L\}} \ket{sl}\bra{sl}$, for both of which the asymptotic state is the (pure) Bell state $\ket{\Phi^+}$.

\smallskip

\emph{Asymptotic behavior.} We now characterize the nature of asymptotic states in terms of the kinematical parameters $\theta$ and $\mu$. In particular, we investigate whether they exhibit maximal entanglement in physically relevant cases (the most general situation is discussed in Appendix~\ref{app:dynGen}).

For Bhabha and M$\o{}$ller scattering, the matrix \eqref{genmappar} takes the following forms, respectively:
\be
    \hspace{-4mm}
    \textbf{M}_{B}=\!\left(\!
    \begin{array}{cccc}
        \! A & \! B & \!  B & \! D \\
        \!-B & \!E & \!F & \! B \\
        \!-B & \!F & \! E & \! B \\
        \!D & \!-B & \! -B &\! A
    \end{array}
    \!\right), \;\,
    \textbf{M}_{M}=\!\left(\!
    \begin{array}{cccc}
        \!A & \! B &\! -B &\! D \\
        \!- B &\! E &\! F &\! -B \\
        \! B &\! F & \!E &\! B \\
        \!D &\! B & \! -B &\! A
    \end{array}
    \!\right).
    \label{MBM}
\ee
These matrices are special realizations of the general matrix \eqref{genmappar}, in which only terms $A, B, D, E, F$ survive, due to specific constraints of the two processes.

\begin{table}[t]
    \begin{ruledtabular}
        \begin{tabular}{lccc}
            \textbf{Process}  & \textbf{Initial state} & \textbf{Regime} & \textbf{Asymptotic state} \\
            \colrule
            Bhabha   & $\ket{RL}$ & u.r. & $\ket{\Psi^{+}}$ \\
            Bhabha   & $\ket{RL}$ & n.r. & $\cos (s_1) \ket{\Phi^{-}} + \sin (s_1) \ket{\Psi^{+}}$ \\
            Bhabha   & $\ket{RR}$ & n.r. & $\ket{\Phi^{+}}$ \\
            \colrule
            M\o{}ller & $\ket{RL}$ & u.r. & $\ket{\Psi^{-}}$ \\
            M\o{}ller & $\ket{RR}$ & n.r. & $\ket{\Phi^{-}}$ \\
            M\o{}ller & $\ket{RL}$ & n.r. & $\cos (s_3) \ket{\Phi^{+}} + \sin (s_3) \ket{\Psi^{-}}$ \\
        \end{tabular}
    \end{ruledtabular}
    \caption{\protect\justifying Maximally entangled asymptotic states obtained by iteration in Bhabha and M\o{}ller scattering processes. Here, u.r.\ and n.r.\ indicate the ultrarelativistic and nonrelativistic regimes, and the angles $s_{1}, s_{3}$ are expressed in terms of the scattering amplitudes~\cite{BlasoneSE2}.}
    \label{tab1}
\end{table}

In the case of Bhabha scattering, two eigenvectors of $\mathbf{M}_{B}$ are Bell states: $\ket{\lambda_1}=\ket{\Phi^{+}}$, with eigenvalue $\lambda_1 = A + D$, and $\ket{\lambda_2} = -\ket{\Psi^{-}}$, with eigenvalue $\lambda_2 = E - F$. The remaining two (not normalized) eigenvectors $\ket{\lambda_3}$ and $\ket{\lambda_4}$ take the form
\begin{eqnarray}\label{NMEE}
    \ket{\lambda_{3,4}} &=& - \ket{\Phi^{-}} - (4 B)^{- 1} \, (a \mp \sqrt{\Delta}) \, \ket{\Psi^{+}}, \\
    \Delta &\equiv & a^2-16B^2, \label{Delta}
\end{eqnarray}
with $a \equiv D + E + F - A$. They are associated with the eigenvalues $\lambda_{3,4} = \frac{1}{2}(b\mp\sqrt{\Delta})
$, where $b \equiv A + E + F - D$. The sign of $\Delta$ defines different cases: for $\Delta > 0$, the eigenvectors $\ket{\lambda_3}, \ket{\lambda_4}$ are maximally entangled states; for $\Delta < 0$, the eigenvectors $\ket{\lambda_3}, \ket{\lambda_4}$ are in general not maximally entangled; for $\Delta = 0$, the eigenvectors $\ket{\lambda_3}, \ket{\lambda_4}$ coalesce\footnote{For the derivation of the spectral properties for Bhabha scattering, see Appendix~\ref{app:bhabhaSpectrum}.}. The corresponding regions in the plane $(\theta, \mu)$ are illustrated in Fig. \ref{fig:degenerate-regions}. Since experiments on Bhabha and M\o{}ller scattering span a massive energy regime -- from a few hundred keV up to the GeV scale \cite{Wimmersperg1987, Scherdin1991, Goebel1993, Acciarri1998} -- we present a logarithmic-scale plot covering the interval from the non-relativistic regime ($\mu\lesssim 1$) to the relativistic one ($\mu=1000$).

\smallskip

As shown  above, the spectral structure of $\mathbf{M}_{B}$ determines the behavior at large $n$. One can verify that $|\lambda_1| > \max\{|\lambda_2|, |\lambda_3|, |\lambda_4|\}$ for all values of $\theta$ and $\mu$. Therefore, if $c_1 \neq 0$ in the expansions (\ref{ISEMP}) or (\ref{NitMS}), iteration of the map \eqref{PMFS2} always leads to the Bell state $\ket{\Phi^+}$, and thus to maximum entanglement, for any initial state. If, at variance, $c_1 = 0$, iteration of the map \eqref{PMFS2} does not generally lead to a maximally entangled state. The situation is described in detail in Fig. \ref{fig:degenerate-regions}. As shown, with increasing energy the asymptotic maximal entanglement overwhelmingly prevails\footnote{Detailed saturation conditions, including modulus-degenerate and exceptional-point regimes, are given in Appendix~\ref{app:iterBhabha}.}.

\smallskip

A similar analysis can be performed for the case of M\o{}ller scattering, where the relevant eigenvectors are $\ket{\Phi^{-}}, \ket{\Psi^{+}}$, and linear combinations of $\ket{\Phi^{+}}, \ket{\Psi^{-}}$.

\smallskip

\emph{QED scattering processes involving fermions and photons}.
When both fermions and photons, and thus different statistics, are involved, the matrices $\textbf{M}$ take the forms provided in Ref.~\cite{NoiChaos}. Their spectral properties restrict the conservation of maximal entanglement, which is guaranteed only for a subset of maximal entangled states in the case of $e^+ e^- \rightarrow \gamma \gamma$, and never for the Compton process. We limit our iteration procedure to the Compton process only; indeed, such a procedure cannot be physically defined for pair annihilation into two photons, except in a purely formal sense\footnote{The pair annihilation into two photons may nevertheless be of interest for quantum optics as a source of entangled photon pairs.}.

For the Compton scattering, map iterations do not converge to fixed points. In the ultrarelativistic regime, successive iterations leave an arbitrary initial state unchanged. By contrast, in the non-relativistic regime, the dynamics exhibits oscillatory behavior. For small values of  $\mu$, concurrence wildly oscillates and entanglement attains appreciable values, although non-maximal. At variance, for $\mu\simeq 10$ the oscillation period becomes large and the entanglement approaches its maximum value.

\smallskip

\emph{Conclusions}.
We have investigated entanglement dynamics in the helicity degrees of freedom in tree-level QED scattering processes at fixed momenta. We introduced a general effective framework for characterizing the dynamics of quantum correlations by modeling scattering processes as quantum maps. Within this approach, the spectral properties of the matrices implementing the maps from the initial to the post-measurement state completely characterize the entanglement dynamics. In particular, these properties guarantee conservation of maximal entanglement in fermion--to--fermion scattering processes. Furthermore, we have shown that the quantum map approach accounts for relevant aspects of the entanglement dynamics (partial conservation of entanglement or asymptotic oscillations) also when both fermion and photons  are involved in scattering. 

Iterating the maps on arbitrary initial states leads to asymptotic fixed points, corresponding to states invariant under further scatterings. For  processes involving fermions only, the convergence to maximally entangled states is  achieved for almost all values of the kinematical parameters $\theta$ and $\mu$. As energy increases, the prevalent convergence to maximal entanglement may suggest a form of ``asymptotic quantum irreversibility" which dominates any intermediate entanglement reduction.  When photons are involved, the presence of particles obeying different statistics  modifies the structure and spectral properties of the corresponding maps, giving rise to  different asymptotic dynamics. 

Our results highlight a strong connection between QED interaction and maximal entanglement. This relation must be traced back to the invariance of the theory under parity  transformations, which enforces specific relations among scattering amplitudes \cite{JacobWick} and, in turn, governs the entanglement dynamics. Maximal entanglement conservation appears thus to be sensitive to the symmetry content of the interaction, and could be used in other contexts as a probe for new physics.

A natural extension of the present work is to investigate whether similar methods provide equally robust insights for other fundamental interactions, given the fact that our results have more general validity beyond QED. Also, going beyond the tree-level approximation by including one-loop contributions~\cite{Kamila} would reveal whether higher-order corrections do affect the above results.

\vspace{0.3cm}

\emph{Acknowledgements}.
We thank Vasco Cavina, Claudio Corian\`o, Stefano Cusumano, and Saverio Pascazio for fruitful discussions. B.M. and K.S. acknowledge: This research was funded in whole or in part by the Austrian Science Fund (FWF) 10.55776/PAT4559623. For open access purposes, the author has applied a CC BY public copyright license to any author-accepted manuscript version arising from this submission.

\bibliographystyle{apsrev4-1}
\bibliography{main}

\appendix


\section{Parity invariance}
\label{app:parity}

In this Appendix, we derive the general structure imposed by parity invariance on the matrix $\mathbf{M}$ entering the map
\begin{equation}
\label{postselectedmap}
    \widetilde{\rho}^{\,f} = \frac{ \mathbf{M}\rho^{\rm in}\mathbf{M}^{T} }{ \operatorname{Tr}\!\left[ \mathbf{M}\rho^{\rm in}\mathbf{M}^{T} \right] },
\end{equation}
where, for the tree-level QED amplitudes considered in this work, the entries of $\mathbf{M}$ are real, so that $\mathbf{M}^{\dagger} = \mathbf{M}^{T}$. The derivation below follows the Jacob--Wick helicity formalism~\cite{JacobWick} and provides the symmetry foundation for the results presented in the main text.

Following Ref.~\cite{JacobWick}, a two-particle helicity state with fixed total angular momentum $J$ and third component $M$ is written as
\begin{equation}
    \ket{p;JM;\lambda_a\lambda_b} = N\int d\Omega\, D^{J\,*}_{M\lambda}(\phi,\theta,-\phi)\, \ket{p,\theta,\phi;\lambda_a\lambda_b},
\end{equation}
where $D^J_{M\lambda}$ is a Wigner $D$-matrix in the conventions of Appendix A of Ref.~\cite{JacobWick}, $N$ is a normalization constant, and
\begin{equation}
    d\Omega=\sin\theta\,d\theta\,d\phi.
\end{equation}
The variable $p$ denotes the magnitude of the relative momentum in the center-of-mass frame. The helicities of particles $a$ and $b$ are denoted by $\lambda_a$ and $\lambda_b$, respectively, and
\begin{equation}
    \lambda=\lambda_a-\lambda_b.
\end{equation}
For spin-$\tfrac12$ particles, $\lambda_{a,b} = \pm\tfrac{1}{2}$, corresponding to right- and left-helicity states.

Parity reverses the momentum of a helicity state and therefore also reverses its helicity. Jacob and Wick introduce the operator
\begin{equation}\label{opy}
    Y = e^{-i\pi J_y}P
\end{equation}
where $P$ is the parity operator. The rotation by $\pi$ about the $y$-axis restores the reference momentum direction after the parity transformation, while leaving the helicity reversed. On a one-particle helicity state,
\begin{equation}\label{Yoneparticle}
    Y\ket{p,\lambda_1} = \eta_1 (-1)^{s_1-\lambda_1} \ket{p,-\lambda_1},
\end{equation}
where $s_1$ and $\eta_1$ are the spin and intrinsic parity of the particle. The phase in Eq.~\eqref{Yoneparticle} follows from the symmetry properties of the Wigner $d$-functions according to Appendix A of Ref.~\cite{JacobWick}. Equivalently, by using Eq.~\eqref{opy}, the parity operator can be written as 
\begin{equation}
    P = e^{i\pi J_y}Y .
\end{equation}

Let $T^J$ denote the fixed-$J$ transition operator. Imposing parity invariance,
\begin{equation}
    P^{-1}SP=S,
\end{equation}
Jacob and Wick obtain
\begin{equation} \label{parampl}
    \bra{-\lambda_c,-\lambda_d} T^J \ket{-\lambda_a,-\lambda_b} = \eta_g\, \bra{\lambda_c,\lambda_d} T^J \ket{\lambda_a,\lambda_b},
\end{equation}
with
\begin{equation} \label{eta}
    \eta_g = \frac{\eta_c\eta_d}{\eta_a\eta_b} (-1)^{s_c+s_d-s_a-s_b}.
\end{equation}

Thus, $\eta_g$ contains both the intrinsic-parity contribution and the spin-dependent Jacob--Wick phase. For all QED processes considered in this work,
\begin{equation}
    \eta_g = 1,
\end{equation}
since the spin-dependent factor in Eq.~\eqref{eta} is unity for four spin-$\tfrac{1}{2}$ particles, while the ratio of intrinsic parities is also equal to unity for the reactions considered here.

Defining
\begin{equation}
    \mu = \lambda_c-\lambda_d,
\end{equation}
the Wigner $d$-functions obey
\begin{equation} \label{wigfprop}
    d^J_{-\lambda,-\mu}(\theta) = d^J_{\mu\lambda}(\theta) = (-1)^{\lambda-\mu} d^J_{\lambda\mu}(\theta).
\end{equation}
The Jacob--Wick partial-wave expansion may be written as
\begin{align}\label{JWexpansion}
    \nonumber \mathcal{M}_{\lambda_c\lambda_d;\lambda_a\lambda_b} (\theta,\phi) &= \frac{1}{p} \sum_J \left(J+\frac12\right) \bra{\lambda_c,\lambda_d} T^J \ket{\lambda_a,\lambda_b} \\ &\qquad \times e^{i(\lambda-\mu)\phi} d^J_{\lambda\mu}(\theta).
\end{align}
Combining Eqs.~\eqref{parampl}, \eqref{eta}, \eqref{wigfprop}, and \eqref{JWexpansion}, one finds
\begin{align} 
    \nonumber \mathcal{M}_{-\lambda_c,-\lambda_d; -\lambda_a,-\lambda_b}(\theta,\phi) &= \eta_g (-1)^{\lambda-\mu} \mathcal{M}_{\lambda_c,\lambda_d; \lambda_a,\lambda_b}(\theta,-\phi) \\
    &= \eta_g \mathcal{M}_{\lambda_c,\lambda_d; \lambda_a,\lambda_b}(\theta,\pi-\phi). \label{mrel}
\end{align}

Eq.~\eqref{mrel} is the basic parity constraint on the helicity amplitudes. At a fixed choice of scattering plane, it relates amplitudes whose four helicities are simultaneously reversed and reduces the number of independent amplitudes from $16$ to $8$.

For completeness, time-reversal invariance gives an additional independent relation between fixed-$J$ helicity amplitudes~\cite{JacobWick}. In the conventions adopted here, it may be written as
\begin{equation}\label{timerev}
    \mathcal{M}_{\lambda_c, \lambda_d; \lambda_a, \lambda_b}(\theta,\phi) = (-1)^{\lambda-\mu} \mathcal{M}_{\lambda_a\lambda_b;\lambda_c\lambda_d}(\theta,\phi).
\end{equation}
The general matrix structure derived below, however, follows from parity alone and does not require Eq.~\eqref{timerev}.

Using the ordered helicity basis
\begin{equation}
    \bigl\{ \ket{RR},\ket{RL},\ket{LR},\ket{LL} \bigr\},
\end{equation}
and the row-and-column convention adopted in the main text, Eq.~\eqref{mrel} implies that the matrix $\mathbf{M}$ has the universal form
\begin{equation}\label{genmappar}
    \mathbf{M} = \left(
    \begin{array}{cccc}
        A & B & C & D \\
        H & E & F & - I \\
        I & F & E & - H \\
        D & -C & -B & A
    \end{array}
    \right),
\end{equation}
where
\begin{equation}
    \begin{aligned}
    A&=\mathcal{M}_{RR;RR}, & B&=\mathcal{M}_{RR;RL}, \\
    C&=\mathcal{M}_{RR;LR}, & D&=\mathcal{M}_{RR;LL}, \\
    H&=\mathcal{M}_{RL;RR}, & I&=\mathcal{M}_{LR;RR}, \\
    E&=\mathcal{M}_{RL;RL}, & F&=\mathcal{M}_{RL;LR}.
    \end{aligned}
\end{equation}
Parity fixes the pattern of signs and equalities in Eq.~\eqref{genmappar}.

\vspace{0.5cm}

\paragraph{Decomposition in the Bell basis.}

The structure of Eq.~\eqref{genmappar} becomes transparent in the Bell basis
\begin{equation}
\label{ordbell}
    \mathcal{B}_{\rm Bell} = \bigl\{ \ket{\Phi^+}, \ket{\Psi^-}, \ket{\Phi^-}, \ket{\Psi^+} \bigr\},
\end{equation}
where
\begin{align}
\label{bstates}
    \ket{\Phi^\pm} &= \frac{\ket{RR}\pm\ket{LL}}{\sqrt2}, & \ket{\Psi^\pm} &= \frac{\ket{RL}\pm\ket{LR}}{\sqrt2}.
\end{align}
In the computational basis, states in Eqs.~\eqref{bstates} are represented by
\begin{equation} \label{phi+psi-}
    \ket{\Phi^+} = \frac{1}{\sqrt2} \begin{pmatrix} 1\\ 0\\ 0\\ 1 \end{pmatrix},
    \qquad
    \ket{\Psi^-} = \frac{1}{\sqrt2} \begin{pmatrix} 0\\ 1\\ -1\\ 0 \end{pmatrix},
\end{equation}
and
\begin{equation} \label{phi-psi+}
    \ket{\Phi^-} = \frac{1}{\sqrt2} \begin{pmatrix} 1\\ 0\\ 0\\ -1 \end{pmatrix},
    \qquad
    \ket{\Psi^+} = \frac{1}{\sqrt2} \begin{pmatrix} 0\\ 1\\ 1\\ 0 \end{pmatrix}.
\end{equation}

Let $\mathcal{P}_{\rm Bell}$ be the orthogonal matrix whose columns are the four Bell states ordered as in ~\eqref{ordbell}. The transformed amplitude matrix is
\begin{equation} \label{bellchange}
    \mathbf{M}^{(\rm Bell)} = \mathcal{P}_{\rm Bell}^{-1} \mathbf{M} \mathcal{P}_{\rm Bell} = \mathcal{P}_{\rm Bell}^{T} \mathbf{M} \mathcal{P}_{\rm Bell}.
\end{equation}
A direct calculation gives
\begin{equation} \label{MBell}
    \mathbf{M}^{(\rm Bell)} = 
    \begin{pmatrix}
        A+D & B-C & 0 & 0 \\
        H-I & E-F & 0 & 0 \\
        0 & 0 & A-D & B+C \\
        0 & 0 & H+I & E+F
    \end{pmatrix}.
\end{equation}
Thus,
\begin{equation} \label{mbellstructure}
    \mathbf{M}^{(\rm Bell)} =
    \begin{pmatrix}
        R_1 & 0\\
        0 & R_2
    \end{pmatrix},
\end{equation}
with
\begin{equation}
    R_1 =
    \begin{pmatrix}
        A+D & B-C\\
        H-I & E-F
    \end{pmatrix},
    \quad
    R_2 =
    \begin{pmatrix}
        A-D & B+C\\
        H+I & E+F
    \end{pmatrix}.
\end{equation}
Therefore, parity decomposes the dynamics into the two independent sectors
\begin{align}\label{Bellsectors}
    \mathcal{H}_1 &= \operatorname{span} \left\{ \ket{\Phi^+},\ket{\Psi^-} \right\}, \\ \mathcal{H}_2 &= \operatorname{span} \left\{ \ket{\Phi^-},\ket{\Psi^+} \right\}.
\end{align}
Such a structure provides a direct proof of maximal-entanglement conservation for any process described by it, together with the associated entanglement dynamics under iteration.

\vspace{0.5cm}

\paragraph{Spectrum.}
By introducing the quantities
\begin{equation} \label{rsqp}
    \begin{aligned}
        R_+ &= A+D+E-F,\\
        S_+ &= A+D-E+F,\\
        Q_+ &= S_+^2+4(B-C)(H-I),
    \end{aligned}
\end{equation}
and
\begin{equation} \label{rsqm}
    \begin{aligned}
        R_- &= A-D+E+F,\\
        S_- &= -A+D+E+F,\\
        Q_- &= S_-^2+4(B+C)(H+I),
    \end{aligned}
\end{equation}
the eigenvalues of $R_1$ can be written as
\begin{equation} \label{eigvalgenp}
    \lambda^{(+)}_{\pm} = \frac12 \left( R_+\pm\sqrt{Q_+} \right),
\end{equation}
and those of $R_2$ are
\begin{equation} \label{eigvalgenm}
    \lambda^{(-)}_{\pm} = \frac12 \left( R_-\pm\sqrt{Q_-} \right).
\end{equation}

For $B-C \neq 0$, corresponding right eigenvectors of the upper block may be chosen, in the helicity basis, as
\begin{equation} \label{vpm}
    v^{(+)}_{\pm} =
    \begin{pmatrix}
        1\\[2mm]
        -\dfrac{S_+ \mp \sqrt{Q_+}}{2(B-C)}\\[3mm]
        \phantom{-}\dfrac{S_+ \mp \sqrt{Q_+}}{2(B-C)}\\[3mm]
        1
    \end{pmatrix}.
\end{equation}
Similarly, for $B+C \neq 0$, eigenvectors of the lower block may be chosen as
\begin{equation} \label{vmp}
    v^{(-)}_{\pm} =
    \begin{pmatrix}
        -1\\[2mm]
        -\dfrac{S_-\pm\sqrt{Q_-}}{2(B+C)}\\[3mm]
        -\dfrac{S_-\pm\sqrt{Q_-}}{2(B+C)}\\[3mm]
        1
    \end{pmatrix}.
\end{equation}
Overall normalization factors have been omitted.

Equations~\eqref{vpm} and \eqref{vmp} make the Bell-sector content explicit:
\begin{equation} \label{combp}
    v^{(+)}_{\pm} = \alpha^{(+)}_{\pm}\ket{\Phi^+} + \beta^{(+)}_{\pm}\ket{\Psi^-},
\end{equation}
and
\begin{equation} \label{combm}
    v^{(-)}_{\pm} = \alpha^{(-)}_{\pm}\ket{\Phi^-} + \beta^{(-)}_{\pm}\ket{\Psi^+},
\end{equation}
where the coefficients may be normalized as convenient. Thus, every eigendirection belongs entirely to one of the two parity-selected Bell sectors in Eq.~\eqref{Bellsectors}.

\vspace{0.5cm}

\paragraph{Preservation of maximal entanglement.}

The Bell-block decomposition also gives a direct proof that, for scattering processes that involved only $1/2$-spin particle with real amplitudes, the map \eqref{postselectedmap} preserves the set of maximally entangled states. As shown in Ref.~\cite{Lamoureux:2003apz}, any maximally entangled two-qubit state can be written, up to an overall phase, as
\begin{equation} \label{mes}
    \ket{\psi_{\rm ME}} = x_1\ket{\Phi^+} + x_2\ket{\Psi^-} + i x_3\ket{\Phi^-} + i x_4\ket{\Psi^+},
\end{equation} 
where $x_j\in\mathbb{R}$. 

Applying Eq.~\eqref{mbellstructure} to Eq.~\eqref{mes} gives
\begin{align}
    \mathbf{M}^{(\rm Bell)}\ket{\psi_{\rm ME}} &= R_1
    \begin{pmatrix}
        x_1\\
        x_2
    \end{pmatrix}
    + i R_2
    \begin{pmatrix}
        x_3\\
        x_4
    \end{pmatrix} \\
    &= y_1\ket{\Phi^+} + y_2\ket{\Psi^-} + i y_3\ket{\Phi^-} + i y_4\ket{\Psi^+},
\end{align}
with $y_j\in\mathbb{R}$  and the resulting state is again maximally entangled.

\vspace{0.5cm}

\paragraph{Iteration and exceptional points.}

The same block structure controls the entanglement dynamics under iteration of the map. The amplitudes depend on the scattering angle $\theta$ and on the dimensionless kinematic parameter
\begin{equation}
    \mu=\frac{|\mathbf{p}|}{m},
\end{equation}
where $\mathbf{p}$ is the incoming three-momentum, and $m$ is the electron mass. Small $\mu$ corresponds to the nonrelativistic regime, whereas $\mu \gg 1$ corresponds to the ultrarelativistic regime.

The asymptotic behavior of an initial state under iteration  of the map, follows the standard spectral-selection mechanism underlying the \textit{power method}: if the map is diagonalizable and the initial state has a non-vanishing component along the eigendirection associated with the unique eigenvalue of largest modulus, the normalized iterates states are asymptotically aligned with that eigendirection. 

This simple selection mechanism is no longer guaranteed at parameter values where two, or more, dominant eigenvalues have the same modulus. In this case, the asymptotic selection of a single eigendirection may fail, and a more careful analysis is required. 

A further possibility arises when an eigenvalue degeneracy is accompanied by the coalescence of the corresponding eigendirections. For the $2\times2$ Bell blocks, a necessary condition is the vanishing of the corresponding discriminant,
\begin{equation} 
    Q_+=0 \qquad \text{or} \qquad Q_-=0.
\end{equation}
If the repeated-eigenvalue block is not proportional to the identity, it is defective and gives rise to a nontrivial Jordan block. Such degeneracies are known as \textit{exceptional points}\footnote{The terminology originates in Kato's perturbation theory of linear operators~\cite{Kato} and is widely used in the modern literature on non-Hermitian and open-system dynamics (see, for example, Ref.~\cite{Minganti2019}).}. After $n$ iterations, the Jordan-block contribution associated with
the eigenvalue $\lambda$ contains the factor
\begin{equation}\label{epeffect}
    \binom{n}{1}\lambda^{n-1} = n\lambda^{n-1},
\end{equation}
which can become asymptotically dominant. The discriminants $Q_\pm$, together with the moduli of the four eigenvalues, therefore determine the distinct regimes of the iterated entanglement dynamics.


\section{Entanglement dynamics from the general structure}
\label{app:dynGen}

In this Appendix, we analyze the entanglement dynamics generated by repeated applications of the parity-constrained map in Eq.~\eqref{genmappar}. No process-specific relations among the matrix elements are imposed beyond the general structure already derived.

We first consider regular parameter values for which the relevant Bell blocks are diagonalizable and the eigendirections below are well defined. The defective cases $Q_\pm=0$ are treated separately through their Jordan decomposition.

\vspace{0.5cm}

\paragraph{Spectral decomposition of the iterated state.}

Using Eqs.~\eqref{vpm} and \eqref{vmp}, the four eigenvectors may be written, up to normalization, as
\begin{align} \label{vpp}
    \ket{v^{(+)}_+} &= \ket{\Phi^+} + \beta^{(+)}_+ \ket{\Psi^-}, \\
    \ket{v^{(+)}_-} &= \ket{\Phi^+} + \beta^{(+)}_- \ket{\Psi^-}, \\
    \ket{v^{(-)}_+} &= \ket{\Phi^-} + \beta^{(-)}_+ \ket{\Psi^+}, \\    
    \ket{v^{(-)}_-} &= \ket{\Phi^-} + \beta^{(-)}_- \ket{\Psi^+}.
\end{align}
The corresponding coefficients are
\begin{align} \label{bpp}
    \beta^{(+)}_+ &= -\frac{S_+ - \sqrt{Q_+}}{2(B-C)}, \\
    \label{bpm} \beta^{(+)}_- &= -\frac{S_+ + \sqrt{Q_+}}{2(B-C)}, \\
    \label{bmp} \beta^{(-)}_+ &= \frac{S_- + \sqrt{Q_-}}{2(B+C)},\\
    \label{bmm} \beta^{(-)}_- &= \frac{S_- - \sqrt{Q_-}}{2(B+C)}.
\end{align}

Assuming that the four eigenvectors form a basis, an arbitrary initial pure state can be expanded as
\begin{align} \label{instate0}
    \ket{i_0} = \frac{1}{\mathcal N_0} \Big[ &c_1\ket{v^{(+)}_+} + c_2 e^{i\varphi}\ket{v^{(+)}_-} \nonumber \\
    &+ c_3 e^{i\xi}\ket{v^{(-)}_+} + c_4 e^{i\tau}\ket{v^{(-)}_-} \Big],
\end{align}
where $c_j\in\mathbb R$, and $\mathcal N_0$ is the normalization factor. Negative signs in the $c_j$ could equivalently be absorbed into the phases; the present parametrization is retained for convenience.

After $n$ applications of the map \eqref{postselectedmap}, the state becomes
\begin{align}
    \nonumber \ket{i_n} &= \frac{1}{\mathcal N_n} \Big[ c_1\left(\lambda^{(+)}_+\right)^n \ket{v^{(+)}_+} \\
    &\qquad + c_2 e^{i\varphi} \left(\lambda^{(+)}_-\right)^n \ket{v^{(+)}_-} \nonumber \\
    \nonumber &\qquad + c_3 e^{i\xi} \left(\lambda^{(-)}_+\right)^n \ket{v^{(-)}_+} \\
    &\qquad + c_4 e^{i\tau} \left(\lambda^{(-)}_-\right)^n \ket{v^{(-)}_-} \Big].
\end{align}

For $Q_\pm>0$, if one eigenvalue is strictly larger in modulus than all the others, and the initial state has a nonvanishing projection onto the corresponding eigendirection, the normalized iterates converge to that eigendirection. Since in this case the eigendirection is a real superposition inside one of the Bell sectors, the iterated state is maximally entangled.

For $Q_+<0$ and/or $Q_-<0$, or whenever the dominant moduli are degenerate in pairs, triplets, or among all four eigenvalues, several contributions can survive asymptotically. These cases can correspond to regions or curves in parameter space where  $\ket{i_n}$, for $n\to\infty$, does not converge to a maximally entangled state unless the initial state $\ket{i_0}$ is already maximally entangled. Therefore, in what follows we verify this statement by imposing directly the condition $\mathcal{C}(|i_n\rangle\langle i_n|) = 1$.

\vspace{0.5cm}

\paragraph{Concurrence criterion.}

Using Eqs.~\eqref{vpp}--\eqref{bmm}, the iterated state can be written in the Bell basis as
\begin{align}\label{iterated-bell}
    \ket{i_n} = \frac{1}{\mathcal N_n} \Big[ &a'\ket{\Phi^+} + b'\ket{\Psi^-} + d'\ket{\Phi^-} + c'\ket{\Psi^+} \Big],
\end{align}
where
\begin{align}
    \label{a'} a' &= c_1\left(\lambda^{(+)}_+\right)^n + c_2 e^{i\varphi} \left(\lambda^{(+)}_-\right)^n, \\
    \label{b'} b' &= c_1\beta^{(+)}_+ \left(\lambda^{(+)}_+\right)^n + c_2 e^{i\varphi}\beta^{(+)}_- \left(\lambda^{(+)}_-\right)^n, \\
    \label{c'} c' &= c_3 e^{i\xi}\beta^{(-)}_+ \left(\lambda^{(-)}_+\right)^n + c_4 e^{i\tau}\beta^{(-)}_- \left(\lambda^{(-)}_-\right)^n, \\
    \label{d'} d' &= c_3 e^{i\xi} \left(\lambda^{(-)}_+\right)^n + c_4 e^{i\tau} \left(\lambda^{(-)}_-\right)^n.
\end{align}
Equivalently, in the computational basis,
\begin{align}
    \ket{i_n} = \frac{1}{\sqrt{2}\mathcal N_n} \Big[ &(a'+d')\ket{RR} + (a'-d')\ket{LL} \nonumber\\
    &+ (b'+c')\ket{RL} + (c'-b')\ket{LR} \Big].
\end{align}

For a normalized pure two-qubit state
\begin{equation}
    \ket{\psi} = A\ket{RR} + B\ket{RL} + C\ket{LR} + D\ket{LL},
\end{equation}
the concurrence is
\begin{equation}
    \mathcal{C}(|\psi\rangle\langle\psi|) = 2|AD-BC|.
\end{equation}
Substituting the coefficients of $\ket{i_n}$ gives
\begin{align}\label{concform}
    \mathcal{C}(|i_n\rangle\langle i_n|) = \frac{ \left| a'^2+b'^2-c'^2-d'^2 \right| }{ |a'|^2+|b'|^2+|c'|^2+|d'|^2 }.
\end{align}
The bound $\mathcal{C}(|i_n\rangle\langle i_n|) \leq 1$ follows from
\begin{equation}
    \left| a'^2+b'^2-c'^2-d'^2 \right| \leq |a'|^2+|b'|^2+|c'|^2+|d'|^2.
\end{equation}
Equality holds if and only if, up to an irrelevant global phase, the coefficients may be chosen such that
\begin{align}
    \label{condc1c2} a',\,b' &\in \mathbb R, \\
    \label{condc3c4} c',\,d' &\in i\mathbb R .
\end{align}
These are precisely the phased-Bell-basis conditions for maximal entanglement.

We now analyze separately the spectral regimes $Q_\pm>0$, $Q_\pm<0$, and $Q_\pm=0$, which give rise to different asymptotic behaviours. The analysis is performed sector by sector. The $Q_+$ sector determines the conditions under which the coefficients $a'$ and $b'$ satisfy Eq.~\eqref{condc1c2}, whereas the $Q_-$ sector determines the conditions under which $c'$ and $d'$ satisfy Eq.~\eqref{condc3c4}. The full condition for concurrence saturation is obtained by combining the conditions associated with the two sectors that are simultaneously realized in the parameter region
under consideration.

\vspace{0.5cm}

\paragraph{The $\mathcal H_1$ sector: Case $Q_+>0$.}

In this case, the eigenvalues $\lambda^{(+)}_\pm$ and the coefficients $\beta^{(+)}_\pm$ are real, and the conditions in Eqs.~\eqref{a'} and \eqref{b'} are verified for
\begin{equation} \label{etacondQp}
    \varphi = k\pi, \qquad k\in\mathbb Z.
\end{equation}

\vspace{0.5cm}

\paragraph{The $\mathcal H_1$ sector: Case $Q_+<0$.}

In this case, the two eigenvalues form a complex-conjugate pair:
\begin{equation}
    \lambda^{(+)}_- = \overline{\lambda^{(+)}_+},
\end{equation}
and the corresponding eigenvector coefficients satisfy
\begin{equation}
    \beta^{(+)}_- = \overline{\beta^{(+)}_+}.
\end{equation}
We define
\begin{align}
    z_n &= \left(\lambda^{(+)}_+\right)^n, \\
    w_n &= \beta^{(+)}_+ \left(\lambda^{(+)}_+\right)^n .
\end{align}
Then
\begin{align}
    \label{apQpneg} a' &= c_1 z_n + c_2 e^{i\varphi}\overline{z_n}, \\
    \label{bpQpneg} b' &= c_1 w_n + c_2 e^{i\varphi}\overline{w_n}.
\end{align}
Since $c_1, c_2 \in \mathbb{R}$, the condition for which both $a'$ and $b'$ are real is
\begin{equation}\label{condQpneg} 
    c_2 e^{i\varphi} = c_1.
\end{equation}
Equivalently,
\begin{equation}
    c_1 = (-1)^k c_2, \quad \varphi = k\pi, \quad k\in\mathbb Z .
\end{equation}

\vspace{0.5cm}

\paragraph{The $\mathcal H_2$ sector: Case $Q_->0$.}

In this case, the eigenvalues $\lambda^{(-)}_\pm$ and the coefficients $\beta^{(-)}_\pm$ are real. Requiring $c'$ and $d'$ to be purely imaginary gives
\begin{align}
    \nonumber \operatorname{Re}(c') &= c_3\beta^{(-)}_+ \left(\lambda^{(-)}_+\right)^n \cos\xi \\
    \nonumber &\qquad + c_4\beta^{(-)}_- \left(\lambda^{(-)}_-\right)^n \cos\tau \\
    &= 0, \label{RecpQmpos} \\
    \nonumber \operatorname{Re}(d') &= c_3 \left(\lambda^{(-)}_+\right)^n \cos\xi \\
    \nonumber &\qquad + c_4 \left(\lambda^{(-)}_-\right)^n \cos\tau \\
    &= 0. \label{RedpQmpos}
\end{align}
Multiplying Eq.~\eqref{RedpQmpos} by $\beta^{(-)}_-$ and
subtracting it from Eq.~\eqref{RecpQmpos} gives
\begin{equation}
    c_3 \left( \beta^{(-)}_+ - \beta^{(-)}_- \right) \left(\lambda^{(-)}_+\right)^n \cos\xi = 0,
\end{equation}
For generic regular values, with
\begin{equation}
    c_3 \neq 0, \qquad \lambda^{(-)}_+ \neq 0, \qquad \beta^{(-)}_+ \neq \beta^{(-)}_-,
\end{equation}
it follows that
\begin{equation}
    \cos\xi = 0.
\end{equation}
Substitution into Eq.~\eqref{RedpQmpos} then gives, provided the second eigendirection also contributes,
\begin{equation}
    \cos\tau = 0.
\end{equation}
Therefore,
\begin{equation} \label{condQmposBoth}
    \xi = \frac{\pi}{2} + k\pi, \qquad \tau = \frac{\pi}{2} + \ell\pi, \qquad k,\ell\in\mathbb Z .
\end{equation}

\vspace{0.5cm}

\paragraph{The $\mathcal H_2$ sector: Case $Q_-<0$.}

In this case, the two eigenvalues form a complex-conjugate pair:
\begin{equation}
    \lambda^{(-)}_- = \overline{\lambda^{(-)}_+},
\end{equation}
and
\begin{equation} \label{betaconjQmneg}
    \beta^{(-)}_- = \overline{\beta^{(-)}_+}.
\end{equation}
We define
\begin{align} \label{znwnQmneg} 
    z'_n &= \left(\lambda^{(-)}_+\right)^n, \\
    w'_n &= \beta^{(-)}_+ \left(\lambda^{(-)}_+\right)^n .
\end{align}
The coefficients in the second Bell sector are then
\begin{align}
    \label{cpQmneg} c' &= c_3 e^{i\xi} w'_n + c_4 e^{i\tau} \overline{w'_n}, \\
    \label{dpQmneg} d' &= c_3 e^{i\xi} z'_n + c_4 e^{i\tau} \overline{z'_n}.
\end{align}
The requirement that $c'$ and $d'$ be purely imaginary is
equivalent to
\begin{align}\label{imcondQmneg} 
    c'+\overline{c'} &=0, \\
    d'+\overline{d'} &=0 .
\end{align}
Using Eqs.~\eqref{cpQmneg} and~\eqref{dpQmneg}, one obtains
\begin{align}
    \left( c_3e^{i\xi} + c_4e^{-i\tau} \right)w'_n + \left( c_3e^{-i\xi} + c_4e^{i\tau} \right)\overline{w'_n} &=0, \label{syscQmneg} \\
    \left( c_3e^{i\xi} + c_4e^{-i\tau} \right)z'_n + \left( c_3e^{-i\xi} + c_4e^{i\tau} \right)\overline{z'_n} &=0 . \label{sysdQmneg}
\end{align}
Multiplying Eq.~\eqref{sysdQmneg} by $\beta^{(-)}_+$ and subtracting it from Eq.~\eqref{syscQmneg} gives
\begin{equation} \label{redQmneg}
    \left( c_3e^{-i\xi} + c_4e^{i\tau} \right) \left( \overline{\beta^{(-)}_+} - \beta^{(-)}_+ \right) \overline{z'_n} = 0 .
\end{equation}
For regular values satisfying
\begin{equation}
    \beta^{(-)}_+ \neq \overline{\beta^{(-)}_+}, \qquad z'_n\neq0,
\end{equation}
it follows that
\begin{equation} \label{phaseeqQmneg}
    c_3e^{-i\xi} + c_4e^{i\tau} = 0 .
\end{equation}
Equivalently,
\begin{equation} \label{condc3Qm}
    c_3 = -c_4e^{i(\tau+\xi)} .
\end{equation}
Since $c_3,c_4\in\mathbb R$, the phase $e^{i(\tau+\xi)}$ must be real. Hence,
\begin{equation}
    \tau = k\pi-\xi, \qquad k\in\mathbb Z,
\end{equation}
and
\begin{equation} \label{condQmneg}
    c_3 = (-1)^{k+1}c_4, \qquad \tau=k\pi-\xi, \qquad k\in\mathbb Z .
\end{equation}

\vspace{0.5cm}

\paragraph{Cases $Q_+ = 0$ and $Q_- = 0$.}

We now consider parameter values for which one or both discriminants vanish. If $Q_+=0$, the two eigenvalues of $R_1$ coalesce at
\begin{equation}
    \lambda_R^{(+)} = \frac{1}{2}\left(A+D+E-F\right),
\end{equation}
whereas, if $Q_-=0$, the two eigenvalues of $R_2$ coalesce at
\begin{equation}
    \lambda_R^{(-)} = \frac{1}{2}\left(A-D+E+F\right).
\end{equation}
Accordingly, the characteristic polynomial of the corresponding block is
\begin{align}
    p_{R_1}(\lambda) &= \left(\lambda-\lambda_R^{(+)}\right)^2, \\
    p_{R_2}(\lambda) &= \left(\lambda-\lambda_R^{(-)}\right)^2.
\end{align}
so that the repeated eigenvalue has algebraic multiplicity two.

For each repeated-eigenvalue block, define its nilpotent part by
\begin{equation}
    \mathfrak{N}_\pm = R_{1,2} - \lambda_R^{(\pm)}\mathbb{I}_2,
\end{equation}
where the indices $+$ and $-$ refer to $R_1$ and $R_2$, respectively, and $\mathbb{I}_2$ denotes a two-dimensional identity matrix. The condition $Q_\pm = 0$ implies
\begin{equation}
    \mathfrak{N}_\pm^2=0.
\end{equation}
Two distinct cases must then be distinguished. If $\mathfrak{N}_\pm=0$, the block is proportional to the identity and is diagonalizable, with geometric multiplicity two. If instead
\begin{equation}
    \mathfrak{N}_\pm \neq 0,
\end{equation}
then $\operatorname{rank}\mathfrak{N}_\pm=1$, the geometric multiplicity is one, and the block is defective. Only the latter case corresponds to an \textit{exceptional point}.

Let $\ket{v_R^{(\pm)}}$ denote an ordinary eigenvector of a defective block,
\begin{equation}
    \left(\mathbf M-\lambda_R^{(\pm)}\mathbb{I}_4\right) \ket{v_R^{(\pm)}}=0,
\end{equation}
where $\mathbb{I}_4$ denotes a four-dimensional identity matrix, and let $\ket{g_R^{(\pm)}}$ be a generalized eigenvector satisfying the Jordan-chain equation
\begin{equation}
    \left(\mathbf M-\lambda_R^{(\pm)}\mathbb{I}_4\right) \ket{g_R^{(\pm)}} = \ket{v_R^{(\pm)}}.
\end{equation}
The generalized eigenvector is defined only up to the transformation
\begin{equation}
    \ket{g_R^{(\pm)}} \longrightarrow \ket{g_R^{(\pm)}}+ r_\pm\ket{v_R^{(\pm)}},
\end{equation}
which does not alter the Jordan-chain relation.

For $B-C\neq0$, a convenient Jordan chain in the upper Bell sector is
\begin{align}
    \ket{v_R^{(+)}} &= \ket{\Phi^+} + \beta_R^{(+)}\ket{\Psi^-}, & \beta_R^{(+)} &= - \frac{S_+}{2(B-C)}, \\
    \ket{g_{R,0}^{(+)}} &= \frac{1}{B-C}\ket{\Psi^-}.
\end{align}
Indeed,
\begin{equation}
    \left(\mathbf M-\lambda_R^{(+)}\mathbb{I}_4\right) \ket{g_{R,0}^{(+)}} = \ket{v_R^{(+)}}.
\end{equation}
Similarly, for $B+C\neq0$, one may choose in the lower Bell sector
\begin{align}
    \ket{v_R^{(-)}} &= \ket{\Phi^-} + \beta_R^{(-)}\ket{\Psi^+}, & \beta_R^{(-)} &= \frac{S_-}{2(B+C)}, \\
    \ket{g_{R,0}^{(-)}} &= \frac{1}{B+C}\ket{\Psi^+},
\end{align}
for which
\begin{equation}
    \left(\mathbf M-\lambda_R^{(-)}\mathbb{I}_4\right) \ket{g_{R,0}^{(-)}} = \ket{v_R^{(-)}}.
\end{equation}
If $B-C=0$ or $B+C=0$, the Jordan chain must instead be constructed directly from the nonzero matrix element of the corresponding nilpotent block.

A general state supported on one Jordan subspace can be written as
\begin{equation}
    \ket{i_0}_{\pm} = \alpha_\pm\ket{v_R^{(\pm)}} + e^{i\zeta_\pm}\beta_\pm\ket{g_R^{(\pm)}}, \qquad \alpha_\pm,\beta_\pm\in\mathbb R.
\end{equation}
Using
\begin{align}
    \mathbf M^n\ket{v_R^{(\pm)}} &= \left(\lambda_R^{(\pm)}\right)^n \ket{v_R^{(\pm)}}, \\
    \mathbf M^n\ket{g_R^{(\pm)}} &= \left(\lambda_R^{(\pm)}\right)^n \ket{g_R^{(\pm)}} + n\left(\lambda_R^{(\pm)}\right)^{n-1} \ket{v_R^{(\pm)}},
\end{align}
one obtains
\begin{align}
    \ket{i_n}_{\pm} = \frac{1}{\mathcal N_n^{(\pm)}} \Big[ & \left( \alpha_\pm\left(\lambda_R^{(\pm)}\right)^n + e^{i\zeta_\pm}\beta_\pm n\left(\lambda_R^{(\pm)}\right)^{n-1} \right) \ket{v_R^{(\pm)}} \nonumber\\
    &+ e^{i\zeta_\pm}\beta_\pm \left(\lambda_R^{(\pm)}\right)^n \ket{g_R^{(\pm)}} \Big].
\end{align}

Assume $\lambda_R^{(\pm)} \neq 0$ and that the initial state has a nonzero generalized-eigenvector component, $\beta_\pm \neq 0$. The term $n(\lambda_R^{(\pm)})^{n-1}$ then dominates the ordinary contribution with the same exponential scaling. Consequently, within that Jordan subspace the normalized iterates converge in projective Hilbert space to the ordinary eigendirection $\ket{v_R^{(\pm)}}$. Since $\beta_R^{(\pm)}$ is real, this eigendirection is a real linear combination of the two Bell states spanning the corresponding sector and is therefore maximally entangled.

Suppose first that both blocks are defective. If
\begin{equation}
    \left|\lambda_R^{(+)}\right| > \left|\lambda_R^{(-)}\right|,
\end{equation}
and the initial state has a generalized-eigenvector component in the upper block, the upper eigendirection $\ket{v_R^{(+)}}$ is selected asymptotically. The analogous conclusion holds with the two sectors interchanged when
\begin{equation}
    \left|\lambda_R^{(-)}\right| > \left|\lambda_R^{(+)}\right|.
\end{equation}

If both blocks are defective and
\begin{equation}
    \left|\lambda_R^{(+)}\right| = \left|\lambda_R^{(-)}\right|,
\end{equation}
then, provided that both generalized-eigenvector components are nonzero, the two Jordan-enhanced terms survive with the same asymptotic order. Up to normalization, the leading contribution is
\begin{align}
    \ket{i_n}_{\mathrm{lead}} &= X_n e^{i\zeta_+} \left( \ket{\Phi^+} + \beta_R^{(+)}\ket{\Psi^-} \right) \nonumber\\
    &+ Y_n e^{i\zeta_-} \left( \ket{\Phi^-} + \beta_R^{(-)}\ket{\Psi^+} \right),
\end{align}
where $X_n,Y_n\in\mathbb R$. If the two repeated eigenvalues have opposite signs, $X_n/Y_n$ may alternate in sign, and the normalized state may approach a period-two orbit rather than a single eigendirection. In either case, Eq.~\eqref{concform} implies concurrence saturation precisely when
\begin{equation}
    \zeta_+-\zeta_- = \frac{\pi}{2}+k\pi, \qquad k\in\mathbb Z.
\end{equation}

If only one block is defective while the other is diagonalizable, a Jordan-enhanced contribution dominates any ordinary contribution with the same eigenvalue modulus, provided that the generalized-eigenvector component is nonzero and the defective eigenvalue does not vanish. The iteration is then driven to the ordinary eigenvector of the defective block, which is maximally entangled. An ordinary eigenvalue of strictly larger modulus in the regular block, however, remains exponentially dominant over the Jordan contribution.

Up to this point, we have characterized all asymptotic regimes leading to concurrence saturation. If the corresponding phase and coefficient constraints on the initial state in Eq.~\eqref{instate0} are not satisfied, then, in each parameter region determined by the sign or vanishing of $Q_\pm$, different asymptotic behaviors may occur. Depending on the initial state, the concurrence may oscillate, converge to a finite value strictly between zero and one, or decay asymptotically to zero.

A systematic classification of these dynamical regimes lies beyond the scope of the present work and is left for future investigation.


\section{Spectral properties of the Bhabha map}
\label{app:bhabhaSpectrum}

The general framework developed in Appendix~\ref{app:dynGen} applies to tree-level $2\to2$ QED processes involving spin-$\tfrac12$ fermions in both the initial and final states. In particular, Bhabha scattering, M\o{}ller scattering, $e^-\mu^-\to e^-\mu^-$, and $e^-e^+\to\mu^-\mu^+$ all satisfy the parity-constrained helicity-amplitude structure of Eq.~\eqref{genmappar}. In this Appendix, we specialize the general analysis to the Bhabha scattering map. The concurrence-saturation conditions obtained for arbitrary parity-invariant maps reduce, in this case, to explicit constraints on the phases and coefficients of the initial state expressed in terms of the Bhabha scattering amplitudes. This provides a unified spectral interpretation of the concurrence dynamics previously derived by direct calculations in the computational basis.

In the relativistic regime a unique maximally entangled eigendirection dominates over most of the kinematic parameter space, leading to concurrence saturation for a broad class of initial states. In the nonrelativistic regime, dominant-modulus degeneracies occur more frequently, and saturation may fail for specific initial states. These cases are analyzed separately below.

Each process obeys additional relations among the helicity amplitudes, arising from its particle content, charge assignments, masses, discrete symmetries, and the contributing $s$-, $t$-, and $u$-channel diagrams. These relations further reduce the number of independent amplitudes. Their explicit expressions are reported in Appendix A of Ref.~\cite{BlasoneSE2}.

As a representative example, we consider Bhabha scattering,
\begin{equation}
    e^-e^+ \to e^-e^+.
\end{equation}
This process provides a useful setting in which to illustrate how the spectral properties of the matrix $\mathbf{M}$ determine the entanglement dynamics generated by repeated scattering.

For Bhabha scattering process, the matrix $\mathbf{M}$ is
\begin{equation} \label{bhabhamap}
    \textbf{M}_{B} = \left(
    \begin{array}{cccc}
        A & B & B & D\\
        -B & E & F & B\\
        -B & F & E & B\\
        D & -B & -B & A
    \end{array}
    \right).
\end{equation}
Eq.~\eqref{bhabhamap} is obtained from the general parity-constrained matrix~\eqref{genmappar} by imposing the additional Bhabha-scattering relations
\begin{equation} \label{relb}
    B=C, \qquad H=I, \qquad H=-B.
\end{equation}
Transforming Eq.~\eqref{bhabhamap} to the ordered Bell basis $\{ \ket{\Phi^+}, \ket{\Psi^-}, \ket{\Phi^-}, \ket{\Psi^+} \}$ according to Eq.~\eqref{bellchange}, one obtains
\begin{equation} \label{bhabhadiag}
    \mathbf{M}_{B}^{(\rm Bell)} = \left(
    \begin{array}{cccc}
        A+D & 0 & 0 & 0 \\
        0 & E-F & 0 & 0 \\
        0 & 0 & A-D & 2B \\
        0 & 0 & -2B & E+F
    \end{array}
    \right).
\end{equation}

Using the explicit tree-level amplitudes reported in Appendix A of Ref.~\cite{BlasoneSE2}, the entries of Eq.~\eqref{bhabhadiag} are
\begin{align} 
    \label{bhabhaentries1} A+D &= \frac{1}{ \mu^2 } \left(2\mu^2+1\right) \csc^2\left(\frac{\theta}{2}\right) , \\
    \label{bhabhaentries2} E-F &= \frac{1}{ \mu^2 } \csc^2\left(\frac{\theta}{2}\right) \left(2\mu^2\cos\theta+1\right) , \\
    \nonumber A-D &= \frac{ \csc^2\left(\frac{\theta}{2}\right) }{ 2\left(\mu^4+\mu^2\right) } \bigl[ \mu^2 \left( \cos(2\theta)+4\mu^2+5 \right) \\
    \label{bhabhaentries3} & \qquad + 2\cos\theta \bigr], \\
    \label{bhabhaentries4} 2B &= \frac{ 2\cot\left(\frac{\theta}{2}\right) \left(\mu^2\cos\theta+1\right) }{ \mu^2\sqrt{\mu^2+1} }, \\
    \label{bhabhaentries5} E+F &= \frac{ \csc^2\left(\frac{\theta}{2}\right) }{ 2\mu^2 } \left[ \mu^2 \left( \cos(2\theta)+3 \right) + 2\cos\theta \right].
\end{align}
The relations in Eq.~\eqref{relb} imply
\begin{equation} \label{relq}
    Q_- = S_-^2 - 16B^2, \qquad Q_+ = S_+^2 .
\end{equation}
The second relation reflects the fact that the upper Bell block is diagonal. Its two eigenvalues are therefore
\begin{align} 
    \label{l1}
    \lambda_1 &= A+D = \frac{\left(2 \mu ^2+1\right) \csc ^2\left(\frac{\theta }{2}\right)}{\mu ^2}, \\
    \label{l2}
    \lambda_2 &= E-F = \frac{\csc ^2\left(\frac{\theta }{2}\right) \left(2 \mu ^2 \cos (\theta )+1\right)}{\mu ^2},
\end{align}
The corresponding normalized eigenvectors are the Bell states
\begin{align}
    \label{sl1} \ket{\lambda_1} &= \ket{\Phi^+}, \\
    \label{sl2} \ket{\lambda_2} &= -\ket{\Psi^-}. 
\end{align}

The remaining two eigenvalues are those of the lower Bell block. For convenience, we define
\begin{equation} \label{ThetaDeltaDef}
    \Theta := \frac{R_-}{2}, \qquad \Delta := Q_-.
\end{equation}
Then the two eigenvalues can be written as
\begin{align}
    \label{l3} \lambda_3 &= \frac{1}{2} \left( R_- - \sqrt{Q_{-}} \right) = \Theta -\frac{\sqrt{\Delta}}{2}, \\
    \label{l4} \lambda_4 &= \frac{1}{2} \left( R_- + \sqrt{Q_{-}} \right) = \Theta + \frac{\sqrt{\Delta}}{2},
\end{align}
Using the explicit Bhabha amplitudes, one finds
\begin{equation} \label{ThetaExplicit}
    \Theta = -\frac{\cos\theta}{\mu^2+1} + \frac{ \cos\theta\, \csc^2\!\left(\frac{\theta}{2}\right) }{ \mu^2 } + \frac{ \cos^2\theta+3 }{ 1-\cos\theta },
\end{equation}
and
\begin{align} \label{DeltaExplicit}
    \nonumber \Delta &= \frac{4}{ (\mu^2+1)^2 } \Bigl[ \left( \mu^2\cos\theta+\mu^2+1 \right)^2 \\
    &\qquad - \dfrac{ 4(\mu^2+1) \cot^2\!\left(\frac{\theta}{2}\right) \left(\mu^2\cos\theta+1\right)^2 }{ \mu^4 } \Bigr].
\end{align}
For $B \neq 0$, the corresponding eigenvectors may be written as
\begin{align}
    \label{sl3} \ket{\lambda_3} &= \frac{1}{\tilde{\mathcal{N}}_3} \left( -\ket{\Phi^-} + C_3\ket{\Psi^+} \right), \\
    \label{sl4} \ket{\lambda_4} &= \frac{1}{\tilde{\mathcal{N}}_4} \left( -\ket{\Phi^-} + C_4\ket{\Psi^+} \right),
\end{align}
where
\begin{align}
    \label{c3} C_3 &= -\frac{ S_- - \sqrt{\Delta} }{ 4B }, \\
    \label{c4} C_4 &= -\frac{ S_- + \sqrt{\Delta} }{ 4B },
\end{align}
and
\begin{equation}
    \tilde{\mathcal{N}}_3 = \sqrt{1+|C_3|^2}, \qquad \tilde{\mathcal{N}}_4 = \sqrt{1+|C_4|^2}.
\end{equation}
are the normalization factors.

The sign of $\Delta$ determines the qualitative spectral structure of the lower Bell block. For $\Delta > 0$, the eigenvalues $\lambda_3$ and $\lambda_4$ are real and distinct, and the associated eigenvectors are real linear combinations of $\ket{\Phi^-}$ and $\ket{\Psi^+}$. They are therefore maximally entangled. For $\Delta < 0$, the two eigenvalues form a complex-conjugate pair, and the corresponding eigendirections are generally not maximally entangled. Finally, at $\Delta=0$, the two eigenvalues coalesce. If the lower block is not proportional to the identity, its eigendirections also coalesce and the matrix reaches an exceptional point.

Thus, the Bhabha map contains two Bell-state eigenvectors in $\mathcal H_1$, while all nontrivial spectral transitions occur in the sector $\mathcal H_2=\operatorname{span}\{\ket{\Phi^-},\ket{\Psi^+}\}$. The relative moduli of $\lambda_1,\lambda_2,\lambda_3,\lambda_4$, together with the sign of $\Delta$, determine the asymptotic entanglement dynamics under iteration.


\section{Iterated Bhabha map}
\label{app:iterBhabha}

We now apply the general spectral framework developed in Appendix~\ref{app:dynGen} to the iterated Bhabha scattering map and study the generated entanglement dynamics. At fixed kinematic parameters $(\theta,\mu)$, the normalized state after $n$ iterations is
\begin{equation}\label{norm_iterate}
    \ket{i_n} = \frac{ \left(\mathbf M^{(\mathrm{Bell})}_{B}\right)^n\ket{i_0} }{ \left\| \left(\mathbf M^{(\mathrm{Bell})}_{B}\right)^n\ket{i_0} \right\| },
\end{equation}
provided that the denominator is nonzero. Equivalently,
\begin{equation}
    \ket{i_n} = \frac{ \mathbf M^{(\mathrm{Bell})}_{B}\ket{i_{n-1}} }{ \left\| \mathbf M^{(\mathrm{Bell})}_{B}\ket{i_{n-1}} \right\| }.
\end{equation}

Away from exceptional points, let the initial state be expanded in the eigenvector basis of $\mathbf M^{(\mathrm{Bell})}_{B}$ as
\begin{equation}\label{initialspectral}
    \ket{i_0} = \sum_{j=1}^{4}c_j\ket{\lambda_j}.
\end{equation}
The corresponding unnormalized iterate is
\begin{equation}\label{non_norm_iterate}
    \left(\mathbf M^{(\mathrm{Bell})}_{B}\right)^n\ket{i_0} = \sum_{j=1}^{4} c_j\lambda_j^n\ket{\lambda_j}.
\end{equation}
Thus, whenever one eigenvalue has modulus strictly larger than all the others and the initial state has a nonzero component along its eigendirection, the normalized sequence converges in projective Hilbert space to that eigendirection. If two or more dominant eigenvalues have the same modulus, several contributions may instead survive asymptotically. Exceptional points require a separate Jordan analysis.

Throughout this Appendix, the coefficients multiplying explicitly displayed relative phases are chosen real. Overall normalization factors of the eigenvectors may be absorbed into these real coefficients.

\vspace{0.5cm}

\paragraph{Spectral selection and dominant eigendirections.}

The normalized sequence of iterated states defined in Eq.~\eqref{norm_iterate} is governed by the spectral-selection mechanism underlying the \textit{power method} (see Sec.~5.3.1 of Ref.~\cite{Quarteroni}). Let a diagonalizable matrix $A\in\mathbb C^{N\times N}$ have eigenvalues ordered by modulus as 
\begin{equation}
    |\lambda_1| > |\lambda_2| \ge |\lambda_3| \geq \cdots \geq |\lambda_N|.
\end{equation}
If $\{x_j\}_{j=1}^{N}$ are the corresponding right eigenvectors and
\begin{equation}
    q^{(0)} = \sum_{j=1}^{N} \alpha_j x_j , \qquad \alpha_1 \neq 0,
\end{equation}
then
\begin{equation}
    A^n q^{(0)} = \alpha_1\lambda_1^n \left[ x_1 + \sum_{j=2}^{N} \frac{\alpha_j}{\alpha_1} \left( \frac{\lambda_j}{\lambda_1} \right)^n x_j \right],
\end{equation}
for $n\geq 1$.

The terms with $j\geq2$ vanish relative to the first one, so the normalized iterates approach the dominant eigendirection, up to a global phase. If several eigenvalues share the largest modulus, this selection need not occur. At defective eigenvalues, Jordan blocks instead introduce polynomial factors in $n$, as discussed in Appendix~\ref{app:dynGen}.

For the Bhabha map, Eqs.~\eqref{initialspectral} and \eqref{non_norm_iterate} give
\begin{equation}\label{initialstate_bhabha}
    \ket{i_0}=\sum_{j=1}^{4}c_j\ket{\lambda_j},
\end{equation}
and
\begin{equation}\label{nthiter_bhabha}
    \ket{i_n} = \frac{1}{\mathcal{N}_n} \sum_{j=1}^{4}c_j\lambda_j^n\ket{\lambda_j},
\end{equation}
where $\mathcal{N}_n$ ensures unit norm of $\ket{i_n}$.

From Eq.~\eqref{l1}, together with the comparison with
Eqs.~\eqref{l2}--\eqref{l4}, one finds
\begin{equation}\label{lambda1_dominant}
    |\lambda_1| > |\lambda_j|, \qquad j=2,3,4 .
\end{equation}
throughout the kinematic domain under consideration. Therefore, for every initial state with
\begin{equation}
    c_1\neq0,
\end{equation}
the normalized iterates satisfy
\begin{equation}
    \ket{i_n} \longrightarrow \ket{\lambda_1} =
\ket{\Phi^+} , \qquad n\to\infty .
\end{equation}
Hence, every such state is asymptotically driven to concurrence saturation.

Having established the universal dominance of $\ket{\lambda_1} = \ket{\Phi^+}$ whenever $c_1\neq0$, we now restrict attention to the invariant subspace
\begin{equation}
    c_1 = 0,
\end{equation}
and analyze the asymptotic competition among the remaining eigenvalues $\lambda_2$, $\lambda_3$, and $\lambda_4$.

As shown in Fig.~\ref{fig:degenerate-regions}, the kinematic parameter space contains curves along which two or more of the moduli $|\lambda_2|$, $|\lambda_3|$, and $|\lambda_4|$ coincide. These include pairwise modulus-degeneracy curves and a triple point at which all three moduli are equal.

The spectral structure of the lower Bell block in Eq.~\eqref{bhabhadiag} is controlled by the sign of $\Delta$. From Eqs.~\eqref{l3}, \eqref{l4}, \eqref{sl3}, and \eqref{sl4}, one finds that, for $\Delta>0$, the eigenvalues $\lambda_3$ and $\lambda_4$ are real and distinct and their eigendirections may be chosen real. For $\Delta < 0$, they form a complex-conjugate pair, and the corresponding coefficients multiplying $\ket{\Psi^+}$ in their Bell-basis eigendirections are also complex conjugates. At $\Delta = 0$, the eigenvalues coalesce, and the lower block is non-diagonalizable provided it is not proportional to the identity. This is the exceptional-point mechanism discussed in Appendix~\ref{app:dynGen}.

We therefore analyze separately the three cases
\begin{equation}
    \Delta > 0, \qquad \Delta < 0, \qquad \Delta = 0,
\end{equation}
and determine which classes of initial states reach concurrence saturation under iteration.

In each regular modulus-degenerate sector, we restrict the initial state to the eigendirections belonging to the dominant modulus-degenerate subspace. Equivalently, all components along strictly subdominant eigendirections are set to zero. This restriction isolates the sectors in which the ordinary power-method selection of a single eigendirection fails. The distinction between saturation and non-saturation is then determined from the concurrence of the normalized iterated state. We denote by $\mathcal{N}_0$ and $\mathcal{N}_n$ the normalization constants of the initial state and the $n$-th iterate, respectively.

\begin{figure*}[t!]
    \centering
    \includegraphics[width=\textwidth]{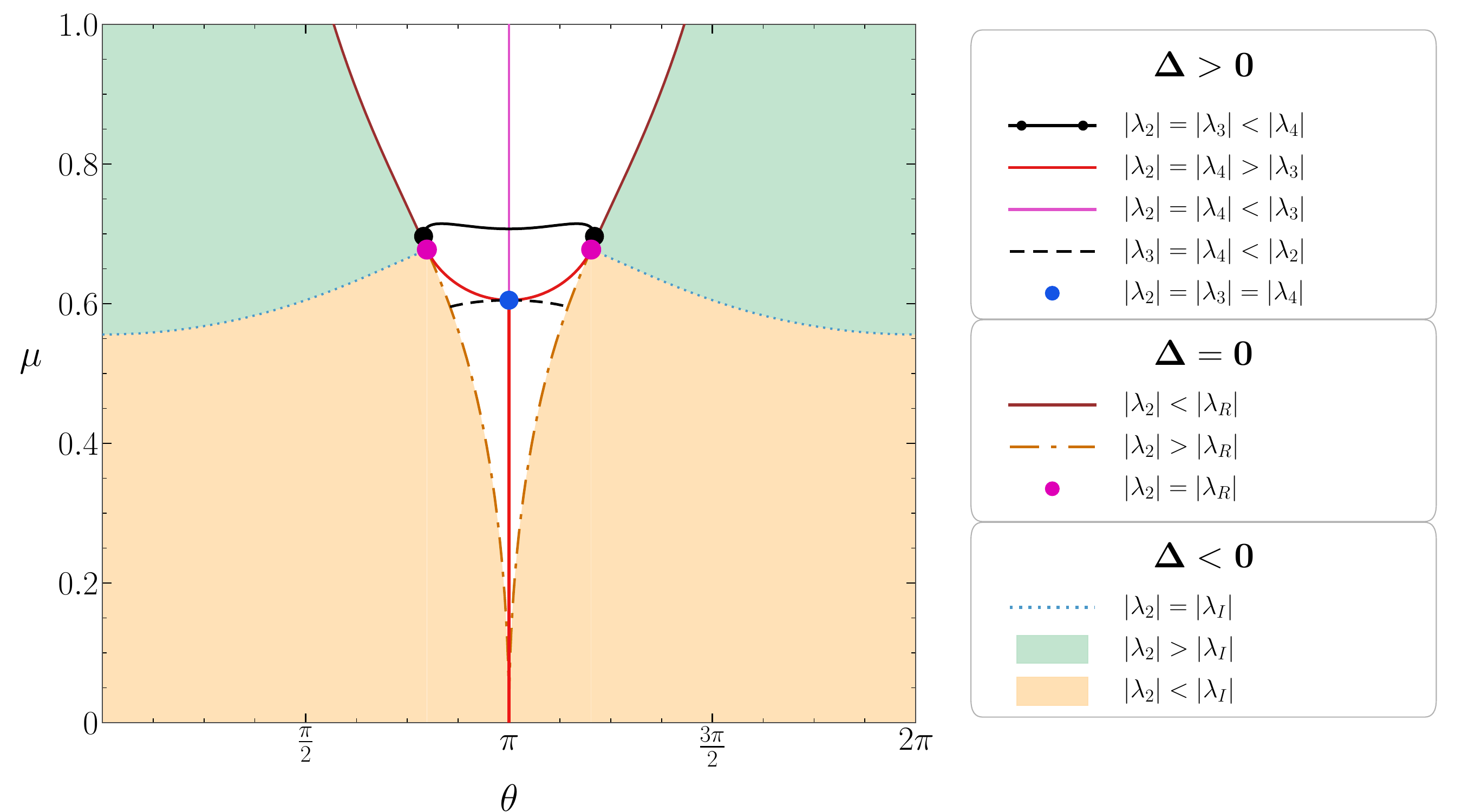}
    \caption{Modulus-degenerate sectors of the Bhabha map in the $(\theta,\mu)$ parameter space.}
    \label{app:fig:degexp}
\end{figure*}

\vspace{0.5cm}

\paragraph{Modulus-degenerate sectors: Case $\Delta > 0$.}

In this region $\lambda_3$ and $\lambda_4$ are real and distinct, and the corresponding eigendirections in Eqs.~\eqref{sl3} and \eqref{sl4} are real. As shown in Fig.~\ref{app:fig:degexp}, one can have pairwise modulus degeneracies among $\lambda_2$, $\lambda_3$, and $\lambda_4$. We first consider the modulus-degeneracy curves in the kinematic parameter space, shown in Fig.~\ref{app:fig:degexp}, and defined by
\begin{equation}\label{deg1}
    |\lambda_2| = |\lambda_j|, \qquad j = 3,4.
\end{equation}
Their analytic expressions are lengthy and not particularly illuminating, and will not be displayed explicitly. Absorbing the normalization constants of the eigenvectors into the real coefficients, a general initial state supported only on such a modulus-degenerate sector can be written as
\begin{equation} \label{init_2j_Deltapos}
    \ket{i_0^{(2,j)}} = \frac{1}{\mathcal{N}_0} \left( c_2\ket{\lambda_2} + e^{i\delta}c_j\ket{\lambda_j} \right), \qquad j = 3,4 ,
\end{equation}
where $c_2, c_j \in \mathbb{R}$. Using Eqs.~\eqref{sl2}--\eqref{sl4}, it can be written in the Bell basis as
\begin{equation}\label{init_2j_expanded}
    \ket{i_0^{(2,j)}} = \frac{1}{\mathcal{N}_0} \left[ -c_2\ket{\Psi^-} + e^{i\delta}c_j \left( -\ket{\Phi^-}  + C_j\ket{\Psi^+} \right) \right],
\end{equation}

After $n$ iterations one obtains
\begin{align} \label{iter_2j_Deltapos}
    \nonumber \ket{i_n^{(2,j)}} &= \frac{1}{\mathcal{N}_n} \Bigl[ -c_2\lambda_2^n\ket{\Psi^-} \\
    &\qquad + e^{i\delta}c_j\lambda_j^n \left( -\ket{\Phi^-} + C_j\ket{\Psi^+} \right) \Bigr].
\end{align}
To determine this state is maximally entangled, we use the concurrence of a pure two-qubit state written in the computational basis. In particular, for a two-qubit state
\begin{equation}\label{twoqubitgen}
    \ket{s} = a\ket{00} + b\ket{01} + c\ket{10} + d\ket{11},
\end{equation}
with normalized coefficients, the concurrence is
\begin{equation} \label{concurrence_formula}
    \mathcal{C}(|s\rangle\langle s|) = 2|ad-bc|.
\end{equation}
For convenience, we absorb the normalization factor in Eq.~\eqref{iter_2j_Deltapos} into the real coefficients and rewrite the state as
\begin{equation}\label{reduced_2j_state}
    \ket{i_n^{(2,j)}} = \alpha_n\ket{\Psi^-} + e^{i\delta} \left( \beta_n\ket{\Phi^-} + \gamma_n\ket{\Psi^+} \right),
\end{equation}
where $\alpha_n,\beta_n,\gamma_n\in\mathbb{R}$, and
\begin{equation}
    \alpha_n^2 + \beta_n^2 + \gamma_n^2 = 1.
\end{equation}
By expressing the Bell states in Eqs.~\eqref{phi+psi-} and \eqref{phi-psi+} in the computational basis, the coefficients of the state in Eq.~\eqref{reduced_2j_state} are
\begin{equation}\label{realcoef}
    \begin{aligned}
        a &= -\frac{e^{i\delta}\beta_n}{\sqrt{2}}, & b &= \frac{-\alpha_n + e^{i\delta}\gamma_n}{\sqrt{2}}, \\
        c &= \frac{\alpha_n + e^{i\delta}\gamma_n}{\sqrt{2}}, & d &= \frac{e^{i\delta}\beta_n}{\sqrt{2}} .
    \end{aligned}
\end{equation}
Therefore, we have
\begin{equation}
    ad-bc = \frac{1}{2} \left[ \alpha_n^2 - e^{2i\delta} \left( \beta_n^2 + \gamma_n^2 \right) \right],
\end{equation}
and the concurrence results
\begin{equation}
    \mathcal{C}(|i_n^{(2,j)}\rangle\langle i_n^{(2,j)}|) = \left| \alpha_n^2 - e^{2i\delta} \left( 1 - \alpha_n^2 \right) \right|.
\end{equation}
Therefore, on the modulus-degeneracy curves defined by Eq.~\eqref{deg1}, a generic initial state of the form Eq.~\eqref{init_2j_Deltapos} does not reach concurrence saturation, except for the relative phases
\begin{equation}
    \delta = \frac{\pi}{2} + k\pi, \qquad k\in\mathbb{Z}.
\end{equation}
For this values of $\delta$, the initial initial state has the form
\begin{equation}\label{inmax_2j_state}
    \ket{i_0^{(2,j)}} = \alpha_0\ket{\Psi^-} + i \left( \beta_0\ket{\Phi^-} + \gamma_0\ket{\Psi^+} \right),
\end{equation}
with $\alpha_0, \beta_0, \gamma_0 \in \mathbb{R}$ and, hence, is a maximal entangled state due to Eq.~\eqref{mes}.

We now consider the modulus-degenerate sector defined by
\begin{equation}\label{deg34_Deltapos}
    |\lambda_3| = |\lambda_4|,
\end{equation}
which corresponds to the violet curve in Fig.~\ref{fig:degenerate-regions}. The general initial state supported only on this modulus-degenerate sector can be written as
\begin{equation}\label{init_34_Deltapos}
    \ket{i_0^{(3,4)}}_{\Delta>0} = \frac{1}{\mathcal{N}_0} \left( c_3\ket{\lambda_3} + e^{i\delta}c_4\ket{\lambda_4} \right),
\end{equation}
where $c_3,c_4\in\mathbb{R}$. After $n$ iterations one obtains
\begin{equation}\label{iter_34_Deltapos}
    \ket{i_n^{(3,4)}}_{\Delta>0} = \frac{1}{\mathcal{N}_n} \left( c_3\lambda_3^n\ket{\lambda_3} + e^{i\delta}c_4\lambda_4^n\ket{\lambda_4} \right).
\end{equation}
Using Eqs.~\eqref{sl3} and \eqref{sl4}, this gives
\begin{equation}
    \begin{aligned}
            \ket{i_n^{(3,4)}}_{\Delta>0} &= \frac{1}{\mathcal{N}_n} \Bigg[ \frac{c_3\lambda_3^n}{\tilde{\mathcal{N}}_3} \left( -\ket{\Phi^-} + C_3\ket{\Psi^+} \right) \\
        &\qquad + e^{i\delta} \frac{c_4\lambda_4^n}{\tilde{\mathcal{N}}_4} \left( -\ket{\Phi^-} + C_4\ket{\Psi^+} \right) \Bigg].
    \end{aligned}
\end{equation}
We now define
\begin{equation} \label{coef_34_Deltapos}
    \begin{aligned}
        c_{3,n} &= \frac{c_3\lambda_3^n}{\tilde{\mathcal{N}}_3}, & c_{4,n} &= \frac{c_4\lambda_4^n}{\tilde{\mathcal{N}}_4}, \\
        C_{3,n} &= \frac{c_3 C_3\lambda_3^n}{\tilde{\mathcal{N}}_3}, & C_{4,n} &= \frac{c_4 C_4\lambda_4^n}{\tilde{\mathcal{N}}_4},
    \end{aligned}
\end{equation}
obtaining
\begin{align}\label{iter_34_reduced_Deltapos}
    \nonumber \ket{i_n^{(3,4)}}_{\Delta>0} &= \frac{1}{\mathcal{N}_n}\Bigl(- \left(c_{3,n} + e^{i\delta}c_{4,n} \right) \ket{\Phi^-} \\
    &\qquad + \left( C_{3,n} + e^{i\delta}C_{4,n} \right) \ket{\Psi^+}\Bigr).
\end{align}
Using Eqs.~\eqref{phi+psi-}, \eqref{phi-psi+}, \eqref{twoqubitgen}, and \eqref{concurrence_formula}, the concurrence of the state~\eqref{iter_34_reduced_Deltapos} is
\begin{equation}\label{conc_34_Deltapos}
    \mathcal{C}(|i_n^{(3,4)} \rangle_{\Delta > 0} \langle i_n^{(3,4)}|) = \frac{ \left|U_n^2 + V_n^2\right|}{|U_n|^2 + |V_n|^2},
\end{equation}
where we have defined
\begin{equation}\label{UV_34_Deltapos}
    \begin{aligned}
        U_n &= c_{3,n} + e^{i\delta}c_{4,n}, \\
        V_n &= C_{3,n} + e^{i\delta}C_{4,n}.
    \end{aligned}
\end{equation}
In order to have $C=1$, the triangle inequality in Eq.~\eqref{conc_34_Deltapos} must be saturated:
\begin{equation}\label{triang_34_Deltapos}
    \left| U_n^2+V_n^2 \right| = |U_n|^2+|V_n|^2 .
\end{equation}
This requires
\begin{equation}
    U_n^2(V_n^*)^2 \in \mathbb{R}_{\geq0},
\end{equation}
or, equivalently,
\begin{equation}\label{condUV_34_Deltapos}
    \operatorname{Im} \left( U_nV_n^* \right) = 0.
\end{equation}
Using Eq.~\eqref{UV_34_Deltapos}, one obtains
\begin{equation}
\label{UVstar_34_Deltapos}
    \begin{aligned}
        U_nV_n^* &= c_{3,n}C_{3,n} + c_{4,n}C_{4,n} \\
        & \qquad + c_{3,n}C_{4,n}e^{-i\delta} + c_{4,n}C_{3,n}e^{i\delta}.
    \end{aligned}
\end{equation}
Since all coefficients are real in the present $\Delta>0$ case, the imaginary part is
\begin{equation} \label{ImUV_34_Deltapos}
    \operatorname{Im} \left( U_nV_n^* \right) = \left( c_{4,n}C_{3,n} - c_{3,n}C_{4,n} \right) \sin\delta .
\end{equation}
From Eq.~\eqref{coef_34_Deltapos}, one has
\begin{equation}
    C_{3,n} = C_3c_{3,n}, \qquad C_{4,n} = C_4c_{4,n}
\end{equation}
and
\begin{equation}
    c_{4,n}C_{3,n} - c_{3,n}C_{4,n} = c_{3,n}c_{4,n} \left( C_3-C_4 \right).
\end{equation}
Since $c_3,c_4 \neq 0$ and $C_3\neq C_4$ is non-vanishing, the condition in Eq.~\eqref{condUV_34_Deltapos} gives
\begin{equation}\label{delta_cond_34_Deltapos}
    \delta = m\pi, \qquad m\in\mathbb{Z}.
\end{equation}
Substituting this condition into the initial state in
Eq.~\eqref{init_34_Deltapos}, one obtains
\begin{equation}\label{init_34_saturating_Deltapos}
    \ket{i_0^{(3,4)}}_{\Delta>0} = \frac{1}{\mathcal{N}_0} \left( c_3\ket{\lambda_3} \pm c_4\ket{\lambda_4} \right),
\end{equation}
where $c_3, c_4\in\mathbb{R}$ Using Eqs.~\eqref{sl3} and \eqref{sl4}, this state is a real linear combination of $\ket{\Phi^-}$ and $\ket{\Psi^+}$. Up to an overall phase, it is therefore a particular case of Eq.~\eqref{mes}, with
\begin{equation}
    x_1 = x_2= 0.
\end{equation}
Hence, also in this modulus-degenerate sector, the initial states that can reach concurrence saturation under the iteration are already maximally entangled.

\vspace{0.5cm}

\paragraph{Modulus-degenerate sectors: The special line $\theta=\pi$.}

We now study the particular case represented by the sectors with $\theta=\pi$. For this value of the scattering angle, the scattering amplitude $B$ vanishes. Therefore, the Bhabha map in Eq.~\eqref{bhabhamap} reduces to 
\begin{equation}\label{bhabhamap_thetapi}  
    \mathbf{M}_{B}^{(\theta=\pi)} = \left( 
    \begin{array}{cccc} 
        A & 0 & 0 & D \\ 
        0 & E & F & 0 \\ 
        0 & F & E & 0 \\ 
        D & 0 & 0 & A 
    \end{array} 
    \right). 
\end{equation}
The corresponding eigenvectors are 
\begin{align} \label{eigvec_thetapi} 
    \ket{\lambda_1} &= \ket{\Phi^+}, \quad \ket{\lambda_2} = -\ket{\Psi^-}, \\ \ket{\lambda_3} &= -\ket{\Phi^-} , \quad \ket{\lambda_4} = \ket{\Psi^+} . 
\end{align}
As before, by using Eqs.~\eqref{phi+psi-} and \eqref{phi-psi+}, one can use the orthogonal change-of-basis matrix $\mathcal{P}_{\rm Bell}$, which diagonalizes the helicity-amplitude map $\mathbf{M}_{B}^{(\theta=\pi)}$ in the Bell basis: 
\begin{equation}\label{bhabhadiag_thetapi}     
    \mathbf{M}_{B}^{{(\rm Bell},\,\theta=\pi)} = \left( 
    \begin{array}{cccc} 
        A+D & 0 & 0 & 0 \\ 
        0 & E-F & 0 & 0 \\ 
        0 & 0 & A-D & 0 \\ 
        0 & 0 & 0 & E+F 
    \end{array}
    \right). 
\end{equation}
Thus, in this case, using
Eqs.~\eqref{bhabhaentries1}--\eqref{bhabhaentries5}, one obtains
\begin{equation}\label{eigval_thetapi}
    \begin{aligned}
        \lambda_1 &= A+D = 2+\frac{1}{\mu^2}, \\
        \lambda_2 &= E-F = -2+\frac{1}{\mu^2}, \\ 
        \lambda_3 &= A-D = 2 + \frac{2}{\mu^2+1} - \frac{1}{\mu^2}, \\
        \lambda_4 &= E+F = 2-\frac{1}{\mu^2}.
    \end{aligned}
\end{equation}

Along the line $\theta=\pi$, one has
\begin{equation}
    |\lambda_2| = |\lambda_4|.
\end{equation}
The comparison with $|\lambda_3|$ splits this line into two different regimes. Introducing the \textit{triple point}
\begin{equation}\label{mustar_thetapi}
    \mu_* = \sqrt{ \frac{1}{2} \left( \sqrt{3}-1 \right) },
\end{equation}
one finds
\begin{equation}
    |\lambda_2| = |\lambda_4| > |\lambda_3|, \qquad 0<\mu<\mu_* ,
\end{equation}
whereas
\begin{equation}
    |\lambda_2| = |\lambda_4| < |\lambda_3|, \qquad \mu>\mu_* .
\end{equation}
At the triple point $\mu=\mu_*$, the three moduli coincide,
\begin{equation}\label{triplepoint_thetapi}
    |\lambda_2| = |\lambda_3| = |\lambda_4|  = 2\mu_{*}^2,
\end{equation}
in particular $\lambda_2 = \lambda_3 = -\lambda_4$.

In the sector defined by $|\lambda_2| = |\lambda_4|$, we consider a general initial state supported only on the corresponding modulus-degenerate eigendirections:
\begin{equation}\label{init_24_thetapi}
    \ket{i_0^{(2,4)}}_{\theta=\pi} = \frac{1}{\mathcal{N}_0} \left( c_2\ket{\lambda_2} + e^{i\delta}c_4\ket{\lambda_4} \right),
\end{equation}
where $c_2, c_4 \in \mathbb{R}$. After $n$ iterations, one obtains
\begin{equation}\label{iter_24_thetapi}
    \ket{i_n^{(2,4)}}_{\theta=\pi} = \frac{1}{\mathcal{N}_n} \left( c_2\lambda_2^n\ket{\lambda_2} + e^{i\delta}c_4\lambda_4^n\ket{\lambda_4} \right).
\end{equation}
Using Eq. ~\eqref{eigval_thetapi}, one has
\begin{equation}
    \lambda_2^n = (-1)^n\lambda_4^n .
\end{equation}
Therefore, absorbing the common factor $\lambda_4^n$ into the normalization constant, the iterated state becomes
\begin{equation}\label{iter_24_reduced_thetapi}
    \ket{i_n^{(2,4)}}_{\theta=\pi} = \frac{1}{\mathcal{N}_n} \left[ - (-1)^n c_2\ket{\Psi^-} + e^{i\delta}c_4\ket{\Psi^+} \right],
\end{equation}
where we used $\mathcal{N}_n^2=c_2^2+c_4^2$. From Eq.~\eqref{concurrence_formula}, one obtains the concurrence of the state~\eqref{iter_24_reduced_thetapi} as
\begin{equation}\label{conc_24_thetapi}
    \mathcal{C}(|i_n^{(2,4)} \rangle_{\theta = \pi} \langle i_n^{(2,4)}|) = \frac{ \left| c_2^2 - e^{2i\delta}c_4^2\right|}{c_2^2+c_4^2}.
\end{equation}
Since $c_2,c_4\in\mathbb{R}$, and assuming $c_2, c_4\neq0$, the condition $\mathcal{C} = 1$ is satisfied only for
\begin{equation}\label{delta_cond_24_thetapi}
    \delta = \frac{\pi}{2} + m\pi, \qquad m\in\mathbb{Z}.
\end{equation}
For these values of the relative phase and using Eq.~\eqref{eigvec_thetapi}, the initial state in Eq.~\eqref{init_24_thetapi} eventually is
\begin{equation}\label{init_24_ME_bell_thetapi}
    \ket{i_0^{(2,4)}}_{\theta=\pi} = \frac{1}{\mathcal{N}_0}\left( -c_2\ket{\Psi^-} \pm i c_4\ket{\Psi^+} \right),
\end{equation}
and up to an overall phase, the state is a particular case of Eq.~\eqref{mes}, with $x_1=x_3=0$.

Therefore, also in this sector, the initial states that can reach concurrence saturation under iteration are already maximally entangled.

We now consider the triple point, defined by
\begin{equation}\label{triple_cond_thetapi}
    |\lambda_2| = |\lambda_3| = |\lambda_4| = \sqrt{3}-1 .
\end{equation}
A general real initial state supported on the corresponding modulus-degenerate sector can be written as
\begin{align}\label{init_234_thetapi}
    \nonumber \ket{i_0^{(2,3,4)}}_{\theta=\pi} &= \frac{1}{\mathcal{N}_0} \Bigl( c_2\ket{\lambda_2} \\
    &\qquad + c_3 e^{i\xi}\ket{\lambda_3} + c_4 e^{i\tau}\ket{\lambda_4} \Bigr),
\end{align}
where $c_2,c_3,c_4\in\mathbb{R}$. At the triple point one has
\begin{equation}
    \lambda_2 = \lambda_3 = -\lambda_4.
\end{equation}
Therefore, after $n$ iterations
\begin{align}\label{iter_234_thetapi}
    \nonumber \ket{i_n^{(2,3,4)}}_{\theta=\pi} &= \frac{1}{\mathcal{N}_n} \Bigl( c_2\lambda_2^n\ket{\lambda_2} + c_3\lambda_3^n e^{i\xi}\ket{\lambda_3} \\
    &\qquad + c_4\lambda_4^n e^{i\tau}\ket{\lambda_4} \Bigr),
\end{align}
eventually one obtains
\begin{align}\label{iter_234_bell_thetapi}
    \nonumber \ket{i_n^{(2,3,4)}}_{\theta=\pi} &= \frac{1}{\mathcal{N}_n} \Bigl( -c_2\ket{\Psi^-} - c_3 e^{i\xi}\ket{\Phi^-} \\
    &\qquad + (-1)^n e^{i\tau}c_4\ket{\Psi^+} \Bigr).
\end{align}
The concurrence then reads
\begin{equation}\label{conc_234_thetapi}
    \mathcal{C}(|i_n^{(2,3,4)} \rangle_{\theta=\pi} \langle i_n^{(2,3,4)}|) = \frac{\left| c_2^2 - e^{2i\xi}c_3^2 - e^{2i\tau}c_4^2 \right| }{ c_2^2 + c_3^2 + c_4^2 }.
\end{equation}
In order to have $\mathcal{C} = 1$, the phases in Eq.~\eqref{conc_234_thetapi} have to saturate the triangle inequality, namely
\begin{equation}\label{triang_234_thetapi}
    \left|c_2^2 - e^{2i\xi}c_3^2 - e^{2i\tau}c_4^2 \right| = c_2^2+c_3^2+c_4^2 .
\end{equation}

In order to verify Eq.~\eqref{triang_234_thetapi}, one must have
\begin{equation}\label{positive_terms_234_thetapi}
    -e^{2i\xi}c_3^2 \in \mathbb{R}_{\geq0}, \qquad -e^{2i\tau}c_4^2 \in \mathbb{R}_{\geq 0}.
\end{equation}
For generic coefficients $c_2,c_3,c_4\neq0$, this gives
\begin{equation}\label{xi_tau_cond_234_thetapi}
    \xi = \frac{\pi}{2} + k\pi, \qquad \tau = \frac{\pi}{2} + \ell\pi, \qquad k,\ell\in\mathbb{Z}.
\end{equation}
Substituting these phases into Eq.~\eqref{init_234_thetapi}, and using Eq.~\eqref{eigvec_thetapi}, one obtains, up to signs that can be absorbed into the real coefficients,
\begin{equation}\label{init_234_ME_thetapi}
    \ket{i_0^{(2,3,4)}}_{\theta=\pi} = \frac{1}{\mathcal{N}_0} \left( -c_2\ket{\Psi^-} + i c_3\ket{\Phi^-} + i c_4\ket{\Psi^+} \right).
\end{equation}
Again, the initial state that saturates the concurrence in this sector is already maximally entangled state.

\vspace{0.5cm}

\paragraph{Modulus-degenerate sectors: Case $\Delta < 0$.}

In this region the eigenvalues $\lambda_3$ and $\lambda_4$ in Eqs. \eqref{l3} and \eqref{l4}  form a complex-conjugate pair, then we define
\begin{equation} \label{modeq}
    |\lambda_I| \equiv |\lambda_3| = |\lambda_4|.
\end{equation}
The corresponding eigendirections are also related by complex conjugation. From Eqs. \eqref{sl3}--\eqref{c4}, since $S_-$ and $B$ are real, one has
\begin{equation}\label{eqc}
    C_4 = C_3^*.
\end{equation}
We set $C_3\equiv C = C_4^*$. Moreover, since the normalization constants depend on $|C_3|$ and $|C_4|$, the relation \eqref{eqc} implies
\begin{equation}
    \tilde{\mathcal{N}}_3 = \tilde{\mathcal{N}}_4 \equiv \tilde{\mathcal{N}}.
\end{equation}
Writing
\begin{equation}
    C = \rho e^{-i\varphi}, \qquad \rho,\varphi\in\mathbb{R},
\end{equation}
the two eigendirections can be written as
\begin{equation}\label{l3m}
    \ket{\lambda_3} = \frac{1}{\tilde{\mathcal{N}}} \left( -\ket{\Phi^-}  + \rho e^{-i\varphi}\ket{\Psi^+} \right),
\end{equation}
and
\begin{equation}\label{l4m}
    \ket{\lambda_4} = \frac{1}{\tilde{\mathcal{N}}} \left( -\ket{\Phi^-}  + \rho e^{i\varphi}\ket{\Psi^+} \right).
\end{equation}
Thus, the eigendirections $\lambda_3$ and $\lambda_4$ contain a non-trivial relative phase between $\ket{\Phi^-}$ and $\ket{\Psi^+}$. In general, these states cannot be brought to the form in Eq.~\eqref{mes} and thus are not, generally speaking, maximally entangled.

We first consider the modulus-degenerate sector defined by Eq.~\eqref{modeq}, with $|\lambda_I|\neq|\lambda_2|$, corresponding to the green and yellow regions in Fig.~\ref{fig:degenerate-regions}. To identify the classes of initial states that can be brought to concurrence saturation under the $n$-times iterated scattering map, we follow the same strategy used for the case $\Delta>0$. We first consider the modulus-degenerate sector defined by Eq.~\eqref{modeq}. The most general initial state supported only on this sector can be written as
\begin{equation}\label{init_34_Deltaneg}
    \ket{i_0^{(3,4)}}_{\Delta<0} = \frac{1}{\mathcal{N}_0} \left( c_3\ket{\lambda_3} + e^{i\delta}c_4\ket{\lambda_4} \right),
\end{equation}
where $c_3, c_4\in\mathbb{R}$. After $n$ iterations one obtains
\begin{equation}\label{iter_34_Deltaneg}
    \ket{i_n^{(3,4)}}_{\Delta<0} = \frac{1}{\mathcal{N}_n} \left( c_3\lambda_3^n\ket{\lambda_3} + e^{i\delta}c_4\lambda_4^n\ket{\lambda_4} \right).
\end{equation}
Since Eq.~\eqref{modeq} holds in this sector, we can write
\begin{equation}\label{l3l4m}
    \lambda_3 = |\lambda_I|e^{-i\omega}, \qquad \lambda_4 = |\lambda_I|e^{i\omega}.
\end{equation}
Using Eqs.~\eqref{l3m} and \eqref{l4m}, and absorbing the common factor $|\lambda_I|^n$ into the normalization constant, the iterated state can be written as
\begin{equation}\label{iter_34_uv}
    \ket{i_n^{(3,4)}}_{\Delta<0} = \frac{1}{\mathcal{N}_n} \left( u_n\ket{\Phi^-} + v_n\ket{\Psi^+} \right),
\end{equation}
where
\begin{equation}\label{uv_Deltaneg}
    \begin{aligned}
        u_n &= - c_3e^{-in\omega} - c_4e^{i(\delta+n\omega)}, \\
        v_n &= \rho \left[ c_3e^{-i(n\omega+\varphi)} + c_4e^{i(\delta+n\omega+\varphi)} \right].
    \end{aligned}
\end{equation}

From Eqs.~\eqref{phi+psi-}, \eqref{phi-psi+},
\eqref{twoqubitgen}, \eqref{concurrence_formula}, \eqref{iter_34_uv} and \eqref{uv_Deltaneg}, the concurrence of the iterated state is
\begin{equation}\label{deltamin0}
    \mathcal{C}(|i_n^{(3,4)} \rangle_{\Delta < 0} \langle i_n^{(3,4)} |) = \frac{ \left| u_n^2+v_n^2 \right|}{|u_n|^2+|v_n|^2}.
\end{equation}
Concurrence saturation, i.e., $\mathcal{C} = 1$, requires the saturation of the triangle inequality
\begin{equation}
    \left| u_n^2+v_n^2 \right| = |u_n|^2 + |v_n|^2.
\end{equation}
This occurs when $u_n^2$ and $v_n^2$ have the same phase, or equivalently when
\begin{equation}\label{conddeltamin}
    \operatorname{Im} \left( u_n v_n^* \right) = 0.
\end{equation}
Using Eq.~\eqref{uv_Deltaneg}, one finds
\begin{equation}
    \begin{aligned}
        u_n v_n^* = -\rho \Big[ & c_3^2 e^{i\varphi} + c_4^2 e^{-i\varphi} \\
        &+ 2c_3c_4 \cos \left( \delta+2n\omega+\varphi \right) \Big].
    \end{aligned}
\end{equation}
The mixed term is real, and therefore
\begin{equation}\label{condfin}
    \operatorname{Im} \left( u_n v_n^* \right) = \rho \left( c_4^2 - c_3^2 \right) \sin\varphi .
\end{equation}
Thus Eq.~\eqref{conddeltamin} gives
\begin{equation}
    \rho \left( c_4^2-c_3^2 \right) \sin\varphi = 0.
\end{equation}
The saturating class of Eq.~\eqref{modeq} is selected by
\begin{equation}\label{condc3c4b}
    c_3 = \pm c_4 ,
\end{equation}
where $c_3,c_4\in\mathbb{R}$.

Substituting the condition in Eq.~\eqref{condc3c4b} into the initial state in Eq.~\eqref{init_34_Deltaneg}, and absorbing the common real coefficient into the normalization constant, one obtains
\begin{equation}\label{pinit}
    \ket{i_0^{(3,4)}}_{\Delta < 0} = \frac{1}{\mathcal{N}_0} \left( \ket{\lambda_3} \pm e^{i\delta}\ket{\lambda_4} \right).
\end{equation}

Using Eqs.~\eqref{eqc}--\eqref{l4m}, the initial state in Eq.~\eqref{pinit} can be written as
\begin{align}\label{rewin}
    \nonumber \ket{i_0^{(3,4)}}_{\Delta<0} &= \frac{1}{\mathcal{N}_0} \Bigl[ - \left( 1\pm e^{i\delta} \right) \ket{\Phi^-} \\
    &\qquad + \rho \left( e^{-i\varphi} \pm e^{i(\delta+\varphi)} \right) \ket{\Psi^+} \Bigr].
\end{align}
For the plus sign, using the identities $1 + e^{i\delta} = 2e^{i\delta/2} \cos(\delta/2)$ and $e^{-i\varphi} + e^{i(\delta+\varphi)} = 2e^{i\delta/2} \cos(\varphi+\delta/2)$, one obtains
\begin{align}\label{plus_sign_state}
    \nonumber \ket{i_0^{(3,4)}}_{\Delta<0} &= \frac{2e^{i\delta/2}}{\mathcal{N}_0} \Bigl[ - \cos\left(\frac{\delta}{2}\right) \ket{\Phi^-} \\
    &\qquad + \rho \cos\left(\varphi+\frac{\delta}{2}\right) \ket{\Psi^+} \Bigr].
\end{align}
For the minus sign, using the identities $1 - e^{i\delta} = -2ie^{i\delta/2} \sin(\delta/2)$ and $e^{-i\varphi} - e^{i(\delta + \varphi)} = -2ie^{i\delta/2} \sin(\varphi + \delta/2)$, one finds
\begin{align}\label{minus_sign_state}
    \nonumber \ket{i_0^{(3,4)}}_{\Delta<0} &= \frac{2ie^{i\delta/2}}{\mathcal{N}_0} \Bigl[ \sin\left(\frac{\delta}{2}\right) \ket{\Phi^-} \\
    &\qquad - \rho \sin\left(\varphi+\frac{\delta}{2}\right) \ket{\Psi^+} \Bigr].
\end{align}
In both cases, up to an overall phase, the coefficients of $\Phi^-$ and $\Psi^+$ are real and then the states in Eqs.~\eqref{plus_sign_state} and \eqref{minus_sign_state} are  maximally entangled.

We now turn to the remaining modulus-degenerate curves in the
$\Delta<0$ region, defined by
\begin{equation}\label{modeq2I} 
    |\lambda_2| = |\lambda_I| .
\end{equation}
These correspond to the two blue curves in Fig.~\ref{fig:degenerate-regions}, which extend toward $\theta=0$ and $\theta=2\pi$, respectively. On these curves,
\begin{equation}
    |\lambda_2| = |\lambda_3| = |\lambda_4|.
\end{equation}

A general initial state supported on
this sector can be written as
\begin{equation}\label{init_2I_Deltaneg}
    \ket{i_0^{(2,I)}} = \frac{1}{\mathcal{N}_0} \left( c_2\ket{\lambda_2} + e^{i\xi}c_3\ket{\lambda_3} + e^{i\tau}c_4\ket{\lambda_4} \right),
\end{equation}
with $c_2$,$c_3$,$c_4$ $\in\mathbb{R}$. After $n$ iterations, one obtains
\begin{align}\label{iter_2I_Deltaneg}
    \nonumber \ket{i_n^{(2,I)}} &= \frac{1}{\mathcal{N}_n} \Bigl( c_2\lambda_2^n\ket{\lambda_2} \\
    &\qquad + e^{i\xi}c_3\lambda_3^n\ket{\lambda_3}+ e^{i\tau}c_4\lambda_4^n\ket{\lambda_4} \Bigr).
\end{align}

Using Eqs.~\eqref{sl2}, \eqref{l3m}, \eqref{l4m}, and \eqref{l3l4m}, together with the fact that, on the sector defined by Eq.~\eqref{modeq2I}, one has
\begin{equation}
    \lambda_2 = \sigma |\lambda_I|, \qquad \sigma = \pm1,
\end{equation}
and it follows that
\begin{equation}
    \lambda_2^n = \sigma_n |\lambda_I|^n, \qquad \sigma_n\equiv\sigma^n=\pm1 .
\end{equation}
Absorbing the common factor $|\lambda_I|^n$ into the normalization constant, and absorbing the common normalization of $\ket{\lambda_3}$ and $\ket{\lambda_4}$ into the real coefficients $c_3$ and $c_4$, we define
\begin{equation}\label{PnQn_Deltaneg}
    \begin{aligned}
        P_n &= c_3 e^{i(\xi-n\omega)} + c_4 e^{i(\tau+n\omega)}, \\
        Q_n &= \rho \left[ c_3 e^{i(\xi-n\omega-\varphi)} + c_4 e^{i(\tau+n\omega+\varphi)} \right].
    \end{aligned}
\end{equation}
The $n$-th iterated state can then be written as
\begin{equation}\label{iter_2I_expanded_Deltaneg}
    \ket{i_n^{(2,I)}} = \frac{1}{\mathcal{N}_n} \left( -c_2\sigma_n\ket{\Psi^-} - P_n\ket{\Phi^-} + Q_n\ket{\Psi^+} \right)
\end{equation}
and the concurrence is
\begin{equation}\label{conc_2I_Deltaneg}
    \mathcal{C}(|i_n^{(2,I)} \rangle \langle i_n^{(2,I)}|) = \frac{ \left| c_2^2 - P_n^2 - Q_n^2 \right|}{c_2^2 + |P_n|^2 + |Q_n|^2}.
\end{equation}
For a generic state with $c_2 \neq 0$, concurrence saturation requires the saturation of a triangle inequality in Eq.~\eqref{conc_2I_Deltaneg}. Since the term $c_2^2$ is real and non-negative, this requires
\begin{equation}
    -P_n^2 = |P_n|^2, \qquad -Q_n^2 = |Q_n|^2 .
\end{equation}
Equivalently,
\begin{equation}
\label{PQimag_2I}
\operatorname{Re}(P_n)=0,
\qquad
\operatorname{Re}(Q_n)=0 .
\end{equation}
Using Eq.~\eqref{PnQn_Deltaneg}, these conditions give
\begin{equation}\label{system_RePQ}
    \begin{aligned}
        & \operatorname{Re} \left[ c_3 e^{i(\xi-n\omega)} + c_4 e^{i(\tau+n\omega)} \right] = 0, \\
        & \operatorname{Re} \left[\rho \left( c_3 e^{i(\xi-n\omega-\varphi)} + c_4 e^{i(\tau+n\omega+\varphi)} \right) \right] = 0 .
    \end{aligned}
\end{equation}
Since $\rho$ is real, Eq.~\eqref{system_RePQ} is equivalent to
\begin{equation}\label{system_cos_2I}
    \begin{aligned}
        & c_3\cos(\xi-n\omega) + c_4\cos(\tau+n\omega) = 0, \\
        & c_3\cos(\xi-n\omega-\varphi) + c_4\cos(\tau+n\omega+\varphi) = 0.
    \end{aligned}
\end{equation}
Letting
\begin{equation}
    A_n=\xi-n\omega, \qquad B_n=\tau+n\omega,
\end{equation}
the second relation in Eq.~\eqref{system_cos_2I} can be written as
\begin{align}\label{cond}
    \nonumber &\cos\varphi \left( c_3\cos A_n + c_4\cos B_n \right) \\
    &\qquad + \sin\varphi \left( c_3\sin A_n - c_4\sin B_n \right) = 0.
\end{align}
Using the  Eq.~\eqref{cond}, and assuming $\sin\varphi\neq0$, one obtains
\begin{equation}
    c_3\sin A_n = c_4\sin B_n .
\end{equation}
Thus the system reduces to
\begin{equation}\label{system_sincos_2I}
    \begin{aligned}
        c_3\cos A_n &= -c_4\cos B_n, \\ 
        c_3\sin A_n &= c_4\sin B_n .
    \end{aligned}
\end{equation}
Equivalently,
\begin{equation}
    c_3e^{iA_n} = -c_4e^{-iB_n}.
\end{equation}
Multiplying by $e^{iB_n}$, and using $A_n + B_n = \xi + \tau$, one finds
\begin{equation}
    e^{i(\xi+\tau)} = -\frac{c_4}{c_3}.
\end{equation}
Therefore, for real $c_3$ and $c_4$, concurrence saturation requires
\begin{equation}
    c_3^2 = c_4^2 .
\end{equation}
More explicitly,
\begin{equation} \label{phase_condition_2I}
    \begin{cases}
        \xi + \tau = (2m+1)\pi, & c_3=c_4, \\
        \xi + \tau = 2m\pi, & c_3=-c_4,
    \end{cases} 
    \qquad m\in\mathbb{Z}.
\end{equation}

We now show that the initial states in Eq.~\eqref{init_2I_Deltaneg} satisfying Eq.~\eqref{phase_condition_2I} are already maximally entangled. Let us first consider
\begin{equation}
    c_3 = c_4\equiv c, \qquad \xi+\tau=(2m+1)\pi .
\end{equation}
Then
\begin{equation}
    e^{i\tau} = e^{i[(2m+1)\pi-\xi]} = -e^{-i\xi},
\end{equation}
and the initial state in Eq.~\eqref{init_2I_Deltaneg} becomes
\begin{equation}
    \ket{i_0^{(2,I)}} = \frac{1}{\mathcal{N}_0} \left( c_2\ket{\lambda_2} + c e^{i\xi}\ket{\lambda_3} - c e^{-i\xi}\ket{\lambda_4} \right).
\end{equation}
The second possibility in Eq.~\eqref{phase_condition_2I}, namely
\begin{equation}
    c_3=-c_4\equiv c, \qquad \xi+\tau=2m\pi ,
\end{equation}
gives the same expression, since in this case $e^{i\tau} = e^{-i\xi}$. Therefore, in both cases, it is sufficient to study
\begin{equation}\label{init_2I_saturating_form}
    \ket{i_0^{(2,I)}} = \frac{1}{\mathcal{N}_0} \left( c_2\ket{\lambda_2} + c e^{i\xi}\ket{\lambda_3} - c e^{-i\xi}\ket{\lambda_4} \right).
\end{equation}
Using Eqs.~\eqref{sl2}, \eqref{l3m}, and \eqref{l4m}, and absorbing the common normalization of $\ket{\lambda_3}$ and $\ket{\lambda_4}$ into $c$, one obtains
\begin{equation}
    \begin{aligned}
        \ket{i_0^{(2,I)}} = \frac{1}{\mathcal{N}_0} \Big[ & -c_2\ket{\Psi^-} \\
        &+ c \left( e^{i\xi} - e^{-i\xi} \right) \left( -\ket{\Phi^-}  \right) \\
        &+ \rho c \left( e^{i(\xi-\varphi)} - e^{-i(\xi-\varphi)} \right) \ket{\Psi^+} \Big].
    \end{aligned}
\end{equation}
Equivalently,
\begin{align}\label{init_2I_ME_form}
    \nonumber \ket{i_0^{(2,I)}} &= \frac{1}{\mathcal{N}_0} \Bigl[ -c_2\ket{\Psi^-} - 2ic\sin\xi\,\ket{\Phi^-} \\
    &\qquad + 2i\rho c\sin(\xi-\varphi)\,\ket{\Psi^+} \Bigr].
\end{align}
Thus the state can be written as
\begin{equation}
    \ket{i_0^{(2,I)}} = x_2\ket{\Psi^-} + i x_3\ket{\Phi^-} + i x_4\ket{\Psi^+},
\end{equation}
where $x_2, x_3, x_4\in\mathbb{R}$, and it is a particular case of the maximally entangled state in Eq.~\eqref{mes} with $x_1 = 0$.

\vspace{0.5cm}

\paragraph{Modulus-degenerate sectors: Case $\Delta = 0$.}

In this region, represented by the orange and bordeaux lines in Fig.~\ref{fig:degenerate-regions}, from Eqs.~\eqref{l3} and \eqref{l4} one can see that $\lambda_3=\lambda_4$. Then, for notational convenience, we define
\begin{equation}\label{lambdaR_def}
    \lambda_R \equiv \lambda_3=\lambda_4 .
\end{equation}
In order to study the spectrum of the lower block in Eq.~\eqref{bhabhadiag}, which, consistently with the notation introduced in Eq.~\eqref{mbellstructure}, we denote by $R_2$, and to which the eigenvalues $\lambda_3$ and $\lambda_4$ refer, we consider the trace and the determinant of $R_2$. Indeed, one obtains
\begin{equation}\label{trace_R2_Delta0}
    \operatorname{Tr}(R_2) = 2\lambda_R,
\end{equation}
and
\begin{equation}\label{det_R2_Delta0}
    \det(R_2) = \lambda_R^2.
\end{equation}
For that $2\times2$ matrix, the Eqs.~\eqref{trace_R2_Delta0} and \eqref{det_R2_Delta0} imply that the characteristic polynomial of $R_2$ is
\begin{equation}\label{charpoly_R2_Delta0_factorized}
    p_{R_2}(x) = (x-\lambda_R)^2 .
\end{equation}
Thus, $\lambda_R$ is a double root of $p_{R_2}$, i.e., it has algebraic multiplicity two.

From Eqs.~\eqref{relb}, \eqref{relq}, and \eqref{sl3}--\eqref{c4}, the eigenvectors $\ket{\lambda_3}$ and $\ket{\lambda_4}$ coalesce along the curves $\Delta=0$. Indeed, Eq.~\eqref{relq} implies
\begin{equation}
    S_- = \pm 4B.
\end{equation}
On the exceptional curve in the interval $0 < \theta < \pi$, where $S_- = -4B$, the coalesced eigenvector is
\begin{equation}\label{lambdaR_eigenvector}
    \ket{\lambda_R^{(+)}} = \frac{1}{2}
    \begin{pmatrix}
        -1\\
        1\\
        1\\
        1
    \end{pmatrix}.
\end{equation}
On the exceptional curve in the interval $\pi < \theta < 2\pi$, where $S_- = 4B$, it is instead
\begin{equation}\label{lambdaRm_eigenvector}
    \ket{\lambda_R^{(-)}} = \frac{1}{2}
    \begin{pmatrix}
        -1\\
        -1\\
        -1\\
        1
    \end{pmatrix}.
\end{equation}
As shown below, these are the ordinary eigenvectors of the nontrivial Jordan blocks associated with the respective exceptional curves. In the following, we present the derivation explicitly for the branch $S_- = -4B$, setting $\ket{\lambda_R} \equiv \ket{\lambda_R^{(+)}}$. The analysis of the opposite branch proceeds analogously and yields the same conclusions.

To determine whether $R_2$ is defective, we calculate the geometric multiplicity of $\lambda_R$, introducing
\begin{equation}\label{JR_def}
    J_R \equiv R_2-\lambda_R I_2 .
\end{equation}
On the same curves, one finds
\begin{equation}\label{JR_det}
    \det J_R = 0,
\end{equation}
while
\begin{equation}\label{JR_nonzero}
    J_R\neq 0.
\end{equation}
Since $J_R$ is a $2\times2$ matrix, Eqs.~\eqref{JR_det}
and \eqref{JR_nonzero} imply
\begin{equation}\label{JR_rank}
    \operatorname{rank}J_R = 1.
\end{equation}
Thus, $\lambda_R$ has geometric multiplicity one.

Since the algebraic multiplicity of $\lambda_R$ is larger than its geometric multiplicity, the block $R_2$ is defective. Hence, on the curves $\Delta = 0$, $R_2$ is represented by a nontrivial Jordan block. These curves therefore correspond to exceptional points, in the sense of spectral degeneracies with coalescing eigenvectors~\cite{Minganti2019}. We shall refer to them as \textit{exceptional curves}.

The ordinary eigenvector associated with the coalesced eigenvalue $\lambda_R$ is the vector $\ket{\lambda_R}$ in Eq.~\eqref{lambdaR_eigenvector}. The corresponding generalized eigenvector is obtained from the Jordan-chain equation
\begin{equation}\label{Jordan_chain_unnormalized}
    \left( \mathbf{M}_{B} - \lambda_R I_4 \right) \ket{\widetilde{\lambda}_G} = \ket{\lambda_R}.
\end{equation}
Solving Eq.~\eqref{Jordan_chain_unnormalized} gives
\begin{equation}\label{lambdaG_tilde_def}
    \ket{\widetilde{\lambda}_G} =
    \begin{pmatrix}
        r\\
        -\frac{1}{4B}-r\\
        -\frac{1}{4B}-r\\
        -r
    \end{pmatrix}, \qquad r\in\mathbb{R}.
\end{equation}
The normalized generalized eigenvector is then
\begin{equation}\label{lambdaG_def}
    \ket{\lambda_G} = \frac{1}{\tilde{\mathcal{N}}_r}
\ket{\widetilde{\lambda}_G},
\end{equation}
with
\begin{equation}\label{Nr_def}
    \tilde{\mathcal{N}}_r = \sqrt{ 2\left[ r^2+ \left( \frac{1}{4B}+r \right)^2 \right] }.
\end{equation}
Using Eqs.~\eqref{lambdaR_eigenvector} and \eqref{lambdaG_def}, a general initial state supported on the Jordan subspace can bewritten as
\begin{equation}\label{init_Jordan_subspace}
    \ket{i_0} = \alpha\ket{\lambda_R} + e^{i\xi}\beta\ket{\lambda_G},
\end{equation}
where $\alpha,\beta\in\mathbb{R}$.

Following the same procedure used in the previous cases, by iterating the Bhabha map in Eq.~\eqref{bhabhamap} on the initial state in Eq.~\eqref{init_Jordan_subspace}, one obtains
\begin{align}\label{iter_Jordan_subspace}
    \nonumber \ket{i_n} &= \frac{1}{\mathcal N_n} \Bigl[ \left( \alpha\lambda_R^n + e^{i\xi}\beta \frac{n}{\tilde{\mathcal{N}}_r} \lambda_R^{n-1} \right) \ket{\lambda_R} \\
    &\qquad + \beta e^{i\xi}\lambda_R^n \ket{\lambda_G} \Bigr].
\end{align}
The $n\lambda_R^{n-1}$ factor dominates the ordinary contribution proportional to $\lambda_R^n$ in the  large-$n$ limit. Hence, the iterated state is asymptotically driven to the eigendirection $\ket{\lambda_R}$ in Eq.~\eqref{lambdaR_eigenvector}. Since $\ket{\lambda_R}$ is maximally entangled, any non-vanishing initial state supported on the Jordan subspace is brought to concurrence saturation under iteration.

The same dominance mechanism applies on the modulus-degenerate sector $|\lambda_2| = |\lambda_R|$, provided $\lambda_R\neq0$. Indeed, when $\beta \neq 0$ in Eq.~\eqref{init_Jordan_subspace}, the Jordan contribution proportional to $n\lambda_R^{n-1}$ dominates, in the limit of large $n$, over the ordinary contributions proportional to $\lambda_R^n$ and $\lambda_2^n$, despite the equality of their moduli. Consequently, the normalized iterated state approaches the eigendirection $\ket{\lambda_R}$, and its concurrence tends to saturation.

An exception arises when $\beta = 0$, namely when the component of the initial state in the Jordan subspace is parallel to the ordinary eigenvector $\ket{\lambda_R}$. In this case, the Jordan-enhanced term vanishes, and the contributions along $\ket{\lambda_R}$ and $\ket{\lambda_2} = -\ket{\Psi^-}$ remain of equal modulus under iteration. If both contributions are nonzero, the normalized state therefore remains a superposition of these two eigendirections, with a relative phase that may depend on $n$. Such a superposition is not necessarily maximally entangled, and its concurrence need not approach unity. Thus, on the modulus-degenerate sector $|\lambda_2|=|\lambda_R|$, asymptotic concurrence saturation holds generically for $\beta\neq0$, while for $\beta=0$ it depends on the relative amplitudes and phases of the two ordinary eigenvector components. Away from this modulus-degenerate sector, the asymptotic behavior is determined by the eigenvalue moduli and the presence or absence of a generalized-eigenvector component, as discussed above.

\end{document}